\documentclass[draftcls, onecolumn, 12 pt]{IEEEtran}

\newcommand{\unicode}[1]{{}}

\usepackage{cite}
\usepackage{graphicx}
\usepackage{epsfig}
\usepackage{amsmath}
\usepackage{amsfonts}
\usepackage{psfrag}
\usepackage{rotating}
\usepackage{latexsym} 
\usepackage{amsthm}
\usepackage{amssymb}
\usepackage{amsmath}
\usepackage{fancyvrb}
\usepackage{textcomp}
\usepackage{url}
\usepackage{ifdraft}
\usepackage{multirow}
\usepackage{rotating}
\usepackage{xspace}
\usepackage{array}
\usepackage[usenames]{color}
\usepackage{arydshln}
\usepackage{multido}
\usepackage{enumerate}
\usepackage[latin1]{inputenc}
\usepackage{color}
\usepackage{xcolor}
\usepackage{soul}
\usepackage{ifsym}
\usepackage{tipa}
\usepackage{graphics}
\usepackage{epsfig}
\usepackage{epstopdf}
\usepackage{amsfonts}
\usepackage{flushend}
\usepackage[caption = false]{subfig}
\usepackage{footmisc}

\usepackage{linegoal} 

\definecolor{SkyBlue}{cmyk}{0.62,0,0.12,0}
\definecolor{BlueGreen}{cmyk}{0.85,0,0.33,0}
\definecolor{Thistle}{cmyk}{0.12,0.59,0,0}

\newbox\HighlightPiece 
\newdimen\FirstHeight
\newdimen\FirstDepth
\newdimen\LastHeight
\newdimen\LastDepth
\newdimen\TotalHeight
\newdimen\TotalDepth
\newdimen\LineIndent
\newdimen\LineGoal

\newif\ifOnePiece

\makeatletter
\newcommand\HLGetDimensions {%
  \setbox\HighlightPiece\lastbox
  \unskip\unpenalty
  \ifdim\wd\HighlightPiece<\linewidth
      \global\OnePiecetrue
  \else
      \global\OnePiecefalse
  \fi 
  \ifOnePiece
  \else
    \global\LastHeight\ht\HighlightPiece
    \global\LastDepth\dp\HighlightPiece
    \loop %
    \setbox\HighlightPiece\lastbox
    \unskip\unpenalty
    \ifvoid\HighlightPiece\else
      \ifdim\wd\HighlightPiece<\linewidth
       \global\FirstHeight\ht\HighlightPiece
       \global\FirstDepth\dp\HighlightPiece
      \fi
    \repeat
 \fi
}
\newcommand\Highlight [2][yellow]
     {%
     \LineGoal\linegoal     
     \LineIndent=\dimexpr\linewidth-\LineGoal\relax
     \vbox{%
     \hfuzz\maxdimen
     \hangindent\LineIndent
     \hangafter\m@ne
     \noindent  #2\endgraf 
     \HLGetDimensions
     }%
     \ifOnePiece
           \begingroup \fboxsep\z@
           \colorbox{#1}{#2}%
           \endgroup
     \else
        \setbox\z@\vbox{%
                        \hfuzz\maxdimen
                        \hangindent\LineIndent
                        \hangafter\m@ne
                        \noindent #2\endgraf }%
        \TotalHeight\ht\z@
        \TotalDepth\dp\z@
        \advance\TotalHeight\TotalDepth
        \advance\TotalHeight-\FirstHeight
        \advance\TotalHeight-\FirstDepth
        \begingroup 
          \color{#1}%
          \rlap{\hbox{\vrule\@height\FirstHeight
                            \@depth\FirstDepth
                            \@width\LineGoal}}%
          \setbox\z@\vbox{\moveleft\LineIndent
                          \hbox{\vrule
                                \@height\TotalHeight
                                \@depth\z@
                                \@width\linewidth}}%
        \ht\z@\z@\dp\z@-\FirstDepth\wd\z@\z@
        \lower\FirstDepth\box\z@
        \endgroup 
         #2%
        \LineGoal\linegoal
        \rlap{\hbox{\begingroup\color{white}%
                        \vrule\@height\LastHeight
                               \@depth\LastDepth
                               \@width\dimexpr\LineGoal+1pt\relax
                    \endgroup}}
     \fi
}%

\makeatother

\def\GatherAndHighlight #1#2\){\Highlight[#1]{$#2$}}

\theoremstyle{plain}

\theoremstyle{plain}

\theoremstyle{plain}

\theoremstyle{plain}

\newtheoremstyle{specialcasestyle}
  {5mm}					
  {5mm}					
  {\upshape}			
  {}					
  {\bfseries\upshape}	
  {.}					
  {1mm}					
  {}					
\theoremstyle{specialcasestyle}

\newcommand{\figref}[1]{Fig.~\protect\ref{#1}}

\newcommand{\tableref}[1]{{Table~\protect\ref{#1}}}

\def\suscript(#1,#2,#3){{#1}^{#2}_{#3}}
\newcommand{\fracparams}[2]{\genfrac{}{}{0pt}{}{{#1}}{{#2}}}

\newcommand{\FoxH}[6][right]{
	\ifthenelse{\equal{#1}{right}}{\suscript(\rm{H},{#2},{#3}){\left[{#4}\left|\fracparams{#5}{#6}\right.\right]}}{
		\ifthenelse{\equal{#1}{left}}{\suscript(\rm{H},{#2},{#3}){\left[\left.{#4}\right|\fracparams{#5}{#6}\right]}}{
			\suscript(\rm{H},{#2},{#3}){\left[{#4}\left|\fracparams{#5}{#6}\right.\right]}
		}
	}
}

\newcommand{\FoxHBar}[6][right]{
	\ifthenelse{\equal{#1}{right}}{\suscript(\rm{\bar{H}},{#2},{#3}){\left[{#4}\left|\fracparams{#5}{#6}\right.\right]}}{
		\ifthenelse{\equal{#1}{left}}{\suscript(\rm{\bar{H}},{#2},{#3}){\left[\left.{#4}\right|\fracparams{#5}{#6}\right]}}{
			\suscript(\rm{\bar{H}},{#2},{#3}){\left[{#4}\left|\fracparams{#5}{#6}\right.\right]}
		}
	}
}

\newcommand{\ExtendedFoxHBar}[6][right]{
	\ifthenelse{\equal{#1}{right}}{\suscript(\rm{\hat{H}},{#2},{#3}){\left[{#4}\left|\fracparams{#5}{#6}\right.\right]}}{
		\ifthenelse{\equal{#1}{left}}{\suscript(\rm{\hat{H}},{#2},{#3}){\left[\left.{#4}\right|\fracparams{#5}{#6}\right]}}{
			\suscript(\rm{\hat{H}},{#2},{#3}){\left[{#4}\left|\fracparams{#5}{#6}\right.\right]}
		}
	}
}

\newcommand{\MeijerG}[6][right]{
	\ifthenelse{\equal{#1}{right}}{\suscript(\rm{G},{#2},{#3}){\left[{#4}\left|\fracparams{#5}{#6}\right.\right]}}{
		\ifthenelse{\equal{#1}{left}}{\suscript(\rm{G},{#2},{#3}){\left[\left.{#4}\right|\fracparams{#5}{#6}\right]}}{
			\suscript(\rm{G},{#2},{#3}){\left[{#4}\left|\fracparams{#5}{#6}\right.\right]}
		}
	}
}
\newcommand{\Hypergeom}[5]{
	\suscript({},{},{#1})\suscript({F},{},{#2})\left[{#3};{#4};{#5} \right]
}

\newcommand{\BesselK}[2][0]{
	\suscript({K},{},{#1})\left({#2}\right)
}

\newcommand{\Expected}[1]{
	{{\mathbb{E}}\left[{#1}\right]}
}

\begin{document}
\title{Performance Analysis of Free-Space Optical Links Over M\'{a}laga ($\mathcal{M}$) Turbulence Channels with Pointing Errors}


\author{Imran~Shafique~Ansari,~\IEEEmembership{Student Member,~IEEE}, Ferkan~Yilmaz,~\IEEEmembership{Member,~IEEE}, and Mohamed-Slim~Alouini,~\IEEEmembership{Fellow,~IEEE}
\normalsize
\thanks{Imran~Shafique~Ansari and Mohamed-Slim~Alouini are with the Computer, Electrical, and Mathematical Sciences and Engineering (CEMSE) Division at King Abdullah University of Science and Technology (KAUST), Thuwal, Makkah Province, Kingdom of Saudi Arabia (e-mail:~\{imran.ansari, slim.alouini\}@kaust.edu.sa).}
\thanks{Ferkan~Yilmaz is currently engaged with Tachyonic Solutions, Turkey (e-mail:~ferkan.yilmaz@alpharabius.solutions; ferkan.yilmaz@tachyonic.solutions).}
\thanks{This work was supported in part by a grant from King Abdulaziz City of Sciences and Technology (KACST) number: AT-34-145.}
\thanks{This work is published in \emph{IEEE Transactions on Wireless Communications}.}
}

\normalsize

\maketitle

\begin{abstract}

In this work, we present a unified performance analysis of a free-space optical (FSO) link that accounts for pointing errors and both types of detection techniques (i.e. intensity modulation/direct detection (IM/DD) as well as heterodyne detection). More specifically, we present unified exact closed-form expressions for the cumulative distribution function, the probability density function, the moment generating function, and the moments of the end-to-end signal-to-noise ratio (SNR) of a single link FSO transmission system, all in terms of the Meijer's G function except for the moments that is in terms of simple elementary functions. We then capitalize on these unified results to offer unified exact closed-form expressions for various performance metrics of FSO link transmission systems, such as, the outage probability, the scintillation index (SI), the average error rate for binary and $M$-ary modulation schemes, and the ergodic capacity (except for IM/DD technique, where we present closed-form lower bound results), all in terms of Meijer's G functions except for the SI that is in terms of simple elementary functions. Additionally, we derive the asymptotic results for all the expressions derived earlier in terms of Meijer's G function in the high SNR regime in terms of simple elementary functions via an asymptotic expansion of the Meijer's G function. We also derive new asymptotic expressions for the ergodic capacity in the low as well as high SNR regimes in terms of simple elementary functions via utilizing moments. All the presented results are verified via computer-based Monte-Carlo simulations.

\end{abstract}


\IEEEpeerreviewmaketitle

\section{Introduction}
\subsection{Background}
\IEEEPARstart{I}{n} recent times, free-space optical (FSO) or optical wireless communication systems have gained an increasing interest due to its advantages including higher bandwidth and higher capacity compared to the traditional RF communication systems. In addition, FSO links are license-free and hence are cost-effective relative to the traditional RF links. It is a promising technology as it offers full-duplex Gigabit Ethernet throughput in certain applications and environment offering a huge license-free spectrum, immunity to interference, and high security \cite{andrews}. These features of FSO communication systems potentially enable solving the issues that the RF communication systems face due to the expensive and scarce spectrum \cite{andrews,peppas1,popoola,park,safari,navidpour,kedar,samimi2,yang4,ansari6}. Additionally, FSO communications does offer bandwidth as the world record stands at 1.2 Tbps or 1200 Gbps \cite{hogan}.
Besides these nice characteristic features of FSO communication systems, they span over long distances of 1Km or longer. However, the atmospheric turbulence may lead to a significant degradation in the performance of the FSO communication systems \cite{andrews}.

Thermal expansion, dynamic wind loads, and weak earthquakes result in the building sway phenomenon that causes vibration of the transmitter beam leading to a misalignment between transmitter and receiver known as pointing error. These pointing errors may lead to significant performance degradation and are a serious issue in urban areas, where the FSO equipments are placed on high-rise buildings \cite{sandalidis2,sandalidis,gappmair}. It is worthy to learn that intensity modulation/direct detection (IM/DD) is the main mode of detection in FSO systems but coherent communications have also been proposed as an alternative detection mode. Among these, heterodyne detection is a more complicated detection method but has the ability to better overcome the thermal noise effects (see \cite{sandalidis,sandalidis2,tsiftsis,liu} and references cited therein).

\subsection{Motivation}
Up until recent past, many irradiance probability density function (PDF) models have been utilized with different degrees of success out of which the most commonly utilized models are the lognormal and the Gamma-Gamma. The scope of lognormal model is restricted to weak turbulences whereas Gamma-Gamma PDF was suggested by Andrews \textit{et. al.} as a reasonable choice because of its much more tractable mathematical model \cite{navas}. Recently, a new and generalized statistical model, M\'{a}laga ($\mathcal{M}$) distribution, was proposed in \cite{navas} to model the irradiance fluctuation of an unbounded optical wavefront (plane or spherical waves) propagating through a turbulent medium under all irradiance conditions in homogeneous, isotropic turbulence \cite{navas1}. This $\mathcal{M}$ distribution unifies most of the proposed statistical models derived until now in the bibliography in a closed-form expression providing an excellent agreement with published simulation data over a wide range of turbulence conditions (weak to strong) \cite{navas}. Hence, both lognormal\,\footnote{The relation between the $\mathcal{M}$ distribution and the lognormal distribution is not an exact relation instead the lognormal distribution is an approximate special case of $\mathcal{M}$ distribution. Similarly, the corresponding relation between the asymptotic results in the two cases are also approximately related to each other (i.e. the asymptotic results applicable for the lornormal distribution as a special case of the $\mathcal{M}$ distribution are actually approximate asymptotic results) rather being exact asymtptotic results.} and Gamma-Gamma models are a special case of this newly proposed general model.

Over the past couple of years, some performance study of a FSO link operating over M\'{a}laga ($\mathcal{M}$) turbulent channels with and without pointing errors in the presence of IM/DD technique has been conducted. Namely, in \cite{navas1}, the authors have derived closed-form expressions for the moments and the error rate of the $\mathcal{M}$ turbulent channel operating under the IM/DD technique in presence of pointing errors whereas in \cite{samimi1}, the authors have derived the error rate in a series form for the coherent differential phase-shift keying in the absence of pointing errors. Additionally, in \cite{balsells,navas2}, the authors have derived the average BER expression for $\mathcal{M}$ distribution under the IM/DD technique with on-off keying (OOK) signaling technique. However as per authors best knowledge, there are no further analysis available in the literature on the $\mathcal{M}$ turbulent FSO channel. Hence, this motivates this work in analyzing and presenting most of the other applicable results as shared in the following subsection.

\subsection{Contributions}
The main contributions of this work are:
\begin{itemize}
\item We present the performance analysis for the $\mathcal{M}$ turbulence channel under the \textit{heterodyne detection} technique in presence of the pointing errors. To the best of our knowledge, these results are new in the literature.
\item Some analysis has been presented in \cite{navas1} for the $\mathcal{M}$ turbulence channel under the \textit{IM/DD technique}. Hence, we complement the work presented in \cite{navas1}. To the best of our knowledge, these complemented results are new in the open literature.
\item Specifically, we derive the PDF, the cumulative distribution function (CDF), and the moment generating function (MGF) of a single $\mathcal{M}$ turbulent FSO link in exact closed-form in terms of Meijer's G function, and the moments in terms of simple elementary functions for both heterodyne and IM/DD detection techniques. Then, we present the outage probability (OP), the bit-error rate (BER) of binary modulation schemes, the symbol error rate (SER) of $M$-ary amplitude modulation (M-AM), $M$-ary phase shift keying (M-PSK) and $M$-ary quadrature amplitude modulation (M-QAM), and the ergodic capacity in terms of Meijer's G functions, and the scintillation index (SI) in terms of simple elementary functions.
\item We derive the \textit{asymptotic expressions} for all the expressions mentioned above in terms of simple elementary functions via Meijer's G function expansion, and additionally, we derive the ergodic capacity at low and high SNR regimes in terms of simple elementary functions by utilizing moments. With the help of these simple results, one can easily derive useful insights. Additionally, these simple results are easily tractable. To the best of our knowledge, these results are new in the open literature.
\item We derive the \textit{diversity order and the coding gain} for $\mathcal{M}$ turbulence model under consideration applicable to both the detection techniques under the presence of pointing error effects. To the best of our knowledge, these results are new in the open literature.
\item Interestingly enough, we are able to unify all the above mentioned results in a unified form i.e. the results for any statistical characteristic or any performance metric applicable to both the detection techniques are presented in a single \textit{unified expression}. To the best of our knowledge, such unified results are new in the open literature. It must be noted here that the study was not straightforward to obtain such novel unified results. After multiple trials, fortunately we were able to derive novel unified expression for the PDF that prove to be simple and straightforward. More importantly, it was not obvious that this unified PDF expression will allow for further analysis but this PDF expression did lead to exact closed-form novel unified expressions for various statistical characteristics and performance metrics.
\item Finally, we also derive the \textit{mapping} between the lognormal distribution parameter and the $\mathcal{M}$ distribution parameters demonstrating the tightness of the approximation of lognormal distribution as a special case of $\mathcal{M}$ distribution.
\end{itemize}

\subsection{Structure}
The remainder of the paper is organized as follows. Sections II and III present a single unified FSO link system and channel model for the $\mathcal{M}$ turbulence distribution accounting for pointing errors with both types of detection techniques (IM/DD and heterodyne) followed by exact closed form expressions and the asymptotic expressions for the statistical characteristics of a single unified FSO link including the CDF and the MGF, and the moments in terms of Meijer's G functions and simple elementary functions, respectively. Subsequently, the performance metrics under consideration, namely, the OP, the SI, the BER, the SER, and the ergodic capacity are also presented in terms of unified expressions and asymptotic expressions in Section IV. Finally, Section V presents some simulation results to validate these analytical results followed by concluding remarks in Section VI.

\section{Channel and System Models}
\subsection{$\mathcal{M}$ Atmospheric Turbulence Model}
The $\mathcal{M}$ turbulence model \cite{navas} is based on a physical model that involves a line-of-sight (LOS) contribution, $U_{L}$, a component that is quasi-forward scattered by the eddies on the propagation axis and coupled to the LOS contribution, $U_{S}^{C}$, and another component, $U_{S}^{G}$, due to energy that is scattered to the receiver by off-axis eddies. $U_{S}^{C}$ and $U_{S}^{G}$ are statistically independent random processes and $U_{L}$ and $U_{S}^{G}$ are also independent random processes. The $\mathcal{M}$ turbulence model can be visually understood via \cite[Fig. 1]{navas}. One of the main motivation to study this turbulence model is its generality i.e. $\mathcal{M}$ represents various other turbulence models as its special case as can be seen from \cite[Table 1]{navas}. Hence, we employ an FSO link that experiences $\mathcal{M}$ turbulence for which the PDF of the irradiance $I_{a}$ is given by \cite[Eq. (24)]{navas}
\begin{equation}\label{Eq:M_PDF}
f_{a}(I_{a})=A\sum_{m=1}^{\beta}a_{m}\,I_{a}\,\BesselK[\alpha-m]{2\sqrt{\frac{\alpha\,\beta\,I_{a}}{g\,\beta+\Omega^{'}}}},\hspace{0.2in}I_{a}>0,
\end{equation}
where
\begin{equation}
\begin{aligned}
A&\triangleq \frac{2\,\alpha^{\alpha/2}}{g^{1+\alpha/2}\Gamma(\alpha)}\left(\frac{g\,\beta}{g\,\beta+\Omega^{'}}\right)^{\beta+\alpha/2},\\
a_{m}&\triangleq \binom{\beta-1}{m-1}\frac{\left(g\,\beta+\Omega^{'}\right)^{1-m/2}}{\left(m-1\right)!}\left(\frac{\Omega^{'}}{g}\right)^{m-1}\left(\frac{\alpha}{\beta}\right)^{m/2},
\end{aligned}
\end{equation}
$\alpha$ is a positive parameter related to the effective number of large-scale cells of the scattering process, $\beta$ is the amount of fading parameter and is a natural number\,\footnote{The expression utilized here is applicable to $\beta$ being a natural number and this allows for this expression to have a finite summation whereas there is a generalized expression of \eqref{Eq:M_PDF} given in \cite[Eq. (22)]{navas} for $\beta$ being a real number though it is less interesting due to the high degree of freedom of the proposed distribution (Sec. III of \cite{navas}) and has an infinite summation.}, $g=\Expected{|U_{S}^{G}|^{2}}=2\,b_{0}\,(1-\rho)$ denotes the average power of the scattering component received by off-axis eddies, $2\,b_{0}=\Expected{|U_{S}^{C}|^{2}+|U_{S}^{G}|^{2}}$ is the average power of the total scatter components, the parameter $0\leq\rho\leq 1$ represents the amount of scattering power coupled to the LOS component, $\Omega^{'}=\Omega+2\,b_{0}\,\rho+2\sqrt{2\,b_{0}\,\rho\,\Omega}\cos(\phi_{A}-\phi_{B})$ represents the average power from the coherent contributions, $\Omega=\Expected{|U_{L}|^{2}}$ is the average power of the LOS component, $\phi_{A}$ and $\phi_{B}$ are the deterministic phases of the LOS and the coupled-to-LOS scatter terms, respectively, $\Gamma(.)$ is the Gamma function as defined in \cite[Eq. (8.310)]{gradshteyn}, and $K_{v}(.)$ is the $v^{\mathrm{th}}$-order modified Bessel function of the second kind \cite[Sec. (8.432)]{gradshteyn}. It is interesting to know here that $\Expected{|U_{S}^{C}|^{2}}=2\,b_{0}\,\rho$ denotes the average power of the coupled-to-LOS scattering component and $\Expected{I_{a}}=\Omega+2\,b_{0}$.\footnote{Detailed information on the $\mathcal{M}$ distribution, its formation, and its random generation can be extracted from \cite[Eqs. (13-21)]{navas}.}

\subsection{Pointing Error Model}
We assume presence of the pointing error impairments for which the PDF of the irradiance $I_{p}$ is given by\,\footnote{For detailed information on the pointing error model and its subsequent derivation, one may refer to \cite{farid}.} \cite[Eq. (11)]{farid}
\begin{equation}\label{Eq:PE}
f_{p}(I_{p})=\frac{\xi^{2}}{A_{0}^{\xi^{2}}}\,I_{p}^{\xi^{2}-1},\hspace{0.3in} 0\leq I_{p}\leq A_{0},
\end{equation}
where $\xi$ is the ratio between the equivalent beam radius at the receiver and the pointing error displacement standard deviation (jitter) at the receiver \cite{sandalidis,gappmair} (i.e. when $\xi \rightarrow \infty$, eq. \eqref{Eq:M_PDF_I} converges to the non-pointing errors case), $A_{0}$ is a constant term that defines the pointing loss given by $A_{0}=\left[\mathrm{erf}\left(v\right)\right]^{2}$, $\mathrm{erf}\left(.\right)$ is the error function \cite[Eq. (7.1.1)]{abramowitz}, $v=\sqrt{\pi}\,a/\left(\sqrt{2}\,w_{z}\right)$, $a$ is the radius of the detection aperture, and $w_{z}$ is the beam waist.

\subsection{Composite Atmospheric Turbulence-Pointing Error Model}
The joint distribution of $I=I_{l}\,I_{a}\,I_{p}$, where $I_{l}$ is the path loss that is a constant in a given weather condition and link distance, can be derived by utilizing
\begin{equation}\label{Eq:Joint_PDF}
\begin{aligned}
f_{I}(I)&=\int_{I/\left(I_{l}\,A_{0}\right)}^{\infty}f_{a}(I_{a})\,f_{I|I_{a}}(I|I_{a})\,dI_{a}\\
&=\int_{I/\left(I_{l}\,A_{0}\right)}^{\infty}f_{a}(I_{a})\,\frac{I}{I_{a}\,I_{l}}f_{p}\left(\frac{I}{I_{a}\,I_{l}}\right)\,dI_{a}.
\end{aligned}
\end{equation}
Hence, applying simple random variable transformation on \eqref{Eq:PE} and using \eqref{Eq:Joint_PDF}, we get the PDF of the receiver irradiance $I$ experiencing $\mathcal{M}$ turbulence in presence of pointing error impairments given by \cite[Eq. (21)]{navas1}
\begin{equation}\label{Eq:M_PDF_I}
f_{I}(I)=\frac{\xi^{2}A}{2\,I}\sum_{m=1}^{\beta}b_{m}\,\MeijerG[right]{3,0}{1,3}{\frac{\alpha\,\beta}{\left(g\,\beta+\Omega^{'}\right)}\frac{I}{I_{l}\,A_{0}}}{\xi^2+1}{\xi^2,\alpha,m},
\end{equation}
where $b_{m}=a_{m}\left[\alpha\,\beta/\left(g\,\beta+\Omega^{'}\right)\right]^{-\left(\alpha+m\right)/2}$ and $\mathrm{G}[.]$ is the Meijer's G function as defined in \cite[Eq. (9.301)]{gradshteyn}.

For the heterodyne detection technique case, the average SNR develops as $\mu_{\mathrm{heterdoyne}}=\eta_{e}\,\mathbb{E}_{I}[I]/N_{0}=I_{l}\,A_{0}\,\eta_{e}\,\xi^{2}(g+\Omega^{'})/\left[\left(1+\xi^{2}\right)N_{0}\right]$,\footnote{$\mathbb{E}_{I}[I^{n}]$ can be easily derived directly utilizing \eqref{Eq:M_PDF_I} though the derived $\mathbb{E}_{I}[I^{n}]$ comes out to be as a summation as expected. Hence to avoid the summation, $\mathbb{E}_{I}[I^{n}]$ has been derived in simpler terms in \cite[Eq. (34)]{navas1} that is utilized here.} where $\eta_{e}$ is the effective photoelectric conversion ratio and $N_{0}$ symbolizes the additive white Gaussian noise (AWGN) sample. Alongside, with $\gamma=\eta_{e}\,I/N_{0}$, we get $I=I_{l}\,A_{0}\,\xi^{2}(g+\Omega^{'})\,\gamma/\left[\mu_{\mathrm{heterodyne}}\,(\xi^{2}+1)\right]$. On utilizing this simple random variable transformation, the resulting SNR PDF under the heterodyne detection technique is given as
\begin{equation}\label{Eq:M_PDF_Heterodyne}
f_{\gamma_{\mathrm{heterodyne}}}(\gamma)=\frac{\xi^{2}A}{2\,\gamma}\sum_{m=1}^{\beta}b_{m}\,\MeijerG[right]{3,0}{1,3}{B\frac{\gamma}{\mu_{\mathrm{heterodyne}}}}{\xi^2+1}{\xi^2,\alpha,m},
\end{equation}
where $B=\xi^{2}\alpha\,\beta\,(g+\Omega^{'})/[\left(\xi^{2}+1\right)(g\,\beta+\Omega^{'})]$ and $\mu_{\mathrm{heterodyne}}=\mathbb{E}_{\gamma_{\mathrm{heterodyne}}}[\gamma]=\overline{\gamma}_{\mathrm{heterodyne}}$ is the average SNR of \eqref{Eq:M_PDF_Heterodyne}.

Similarly, for the IM/DD detection technique case, the electrical SNR develops as $\mu_{\mathrm{IM/DD}}=\eta_{e}^{2}\,\mathbb{E}_{I}^{2}[I]/N_{0}=I_{l}^{2}\,A_{0}^{2}\,\eta_{e}^{2}\,\xi^{4}\left(g+\Omega^{'}\right)^{2}/\left[\left(1+\xi^{2}\right)^{2}N_{0}\right]$. With $\gamma=\eta_{e}^{2}\,I^{2}/N_{0}$, we get $I=\xi^{2}\,(g+\Omega^{'})\,I_{l}\,A_{0}/\left(\xi^{2}+1\right)\sqrt{\gamma/\mu_{\mathrm{IM/DD}}}$. On utilizing this simple random variable transformation, the resulting SNR PDF under the IM/DD technique is given as
\begin{equation}\label{Eq:M_PDF_IMDD}
f_{\gamma_{\mathrm{IM/DD}}}(\gamma)=\frac{\xi^{2}A}{4\,\gamma}\sum_{m=1}^{\beta}b_{m}\,\MeijerG[right]{3,0}{1,3}{B\sqrt{\frac{\gamma}{\mu_{\mathrm{IM/DD}}}}}{\xi^2+1}{\xi^2,\alpha,m},
\end{equation}
where
\begin{equation}\label{Eq:mu_IMDD}
\begin{aligned}
\mu&_{\mathrm{IM/DD}}=\mathbb{E}_{\gamma_{\mathrm{IM/DD}}}[\gamma]\,\mathbb{E}_{I}^{2}[I]/\mathbb{E}_{I}[I^{2}]\\
&=\frac{\xi^{2}\left(\xi^{2}+1\right)^{-2}\left(\xi^{2}+2\right)\left(g+\Omega^{'}\right)}{\alpha^{-1}\left(\alpha+1\right)\left[2\,g\left(g+2\,\Omega^{'}\right)+\Omega^{'^{2}}\left(1+1/\beta\right)\right]}\,\overline{\gamma}_{\mathrm{IM/DD}},
\end{aligned}
\end{equation}
is the electrical SNR of \eqref{Eq:M_PDF_IMDD}. When $\xi^{2}\,(g+\Omega^{'})/\left(\xi^{2}+1\right)=1$, this PDF given in \eqref{Eq:M_PDF_IMDD} comes in agreement with \cite[Eq. (19)]{samimi}.

\subsection{Unification}
Both these PDFs in \eqref{Eq:M_PDF_Heterodyne} and \eqref{Eq:M_PDF_IMDD} can be easily combined yielding the unified expression for the $\mathcal{M}$ turbulence as
\begin{equation}\label{Eq:MPDF}
f_{\gamma}(\gamma)=\frac{\xi^{2}A}{2^{r}\,\gamma}\sum_{m=1}^{\beta}b_{m}\,\MeijerG[right]{3,0}{1,3}{B\left(\frac{\gamma}{\mu_{r}}\right)^{\frac{1}{r}}}{\xi^2+1}{\xi^2,\alpha,m},
\end{equation}
where $r$ is the parameter defining the type of detection technique (i.e. $r=1$ represents heterodyne detection and $r=2$ represents IM/DD). More specifically, for $\mu_{r}$, when $r=1$, $\mu_{1}=\mu_{\mathrm{heterodyne}}$ and when $r=2$, $\mu_{2}=\mu_{\mathrm{IM/DD}}$. Now, as a special case, when $\rho=1$ and $\Omega^{'}=1$ \cite[Table 1]{navas}, this PDF in \eqref{Eq:MPDF} reduces to
\begin{equation}\label{Eq:Gamma_GammaPDF}
f_{\gamma}(\gamma)=\frac{\xi^2}{r\,\gamma\,\Gamma(\alpha)\Gamma(\beta)}\,\MeijerG[right]{3,0}{1,3}{\frac{\xi^2\,\alpha\,\beta}{\xi^2+1}\left(\frac{\gamma}{\mu_{r}}\right)^{\frac{1}{r}}}{\xi^2+1}{\xi^2,\alpha,\beta}.
\end{equation}
This expression in \eqref{Eq:Gamma_GammaPDF} represents the unified PDF for the Gamma-Gamma turbulence and for $\xi^{2}>>1$, eq. \eqref{Eq:Gamma_GammaPDF} reduces to \cite[Eq. (4)]{ansari11}. Additionally, for negligible pointing errors case under IM/DD technique (i.e. $\xi\rightarrow\infty$ and $r=2$) and $\xi^{2}\,>>1$, eq. \eqref{Eq:Gamma_GammaPDF} reduces to \cite[Eq. (9)]{gappmair}.

\subsection{Important Outcomes}
\begin{itemize}
\item It is important to note here that one may easily derive a PDF corresponding to a certain detection technique from the PDF of the other corresponding detection technique via simple random variable transformation. For instance, eq. \eqref{Eq:M_PDF_IMDD} can be easily derived from \eqref{Eq:M_PDF_Heterodyne} by transforming the random variable, $\gamma$, in \eqref{Eq:M_PDF_Heterodyne} to $\gamma^2\,\mu_{\mathrm{IM/DD}}/\mu_{\mathrm{heterodyne}}^{2}$ wherein this updated $\gamma$ will represent the random variable of \eqref{Eq:M_PDF_IMDD}.
\item For readers clarification, there are two different expressions for the two different cases dependent on the type of receiver detection and these differ in various aspects though we would like to share that this unification presented in this work is 'unified' in a rather notational point of view. We classify this unification in terms of having, inclusive within a single expression as in \eqref{Eq:MPDF}, the parameters that characterize the effects of turbulence i.e. $\alpha$ and $\beta$, the parameter that characterizes the effect of pointing errors i.e. $\xi$, the $\mu_{r}$, and the ultimate unifying parameter (notationally speaking) $r$ wherein when we place $r=1$, it gives us the PDF applicable to the heterodyne detection technique with its subsequent $\mu_{1}$ and when we place $r=2$, it gives us the PDF applicable to the IM/DD technique with its subsequent $\mu_{2}$.
\item Emphasizing on the notational importance of this unified expression, we would like to clarify that for $r=1$ case, $\mu_{1}$ represents the average SNR for the heterodyne detection technique whereas for $r=2$ case, $\mu_{2}$ represents the electrical SNR for the IM/DD technique for which its relation with the average SNR is shared in \eqref{Eq:mu_IMDD}.
\end{itemize}

\section{Closed-Form Statistical Characteristics}
\subsection{Cumulative Distribution Function}
\subsubsection{Exact Analysis} Using \cite[Eq. (07.34.21.0084.01)]{mathematica} and some simple algebraic manipulations, the CDF for the $\mathcal{M}$ turbulence can be shown to be given by
\begin{equation}\label{Eq:M_CDF}
F_\gamma(\gamma)=\int_0^{\gamma}f_{\gamma}(t)\,dt=D\sum_{m=1}^{\beta}c_{m}\,\MeijerG[right]{3r,1}{r+1,3r+1}{E\frac{\gamma}{\mu_{r}}}{1,\kappa_{1}}{\kappa_{2},0},
\end{equation}
where $D=\xi^{2}A/\left[2^{r}(2\,\pi)^{r-1}\right]$, $c_{m}=b_{m}\,r^{\alpha+m-1}$, $E=B^{\,r}/r^{2\,r}$, $\kappa_1=\frac{\xi^2+1}{r},\dots,\frac{\xi^2+r}{r}$ comprises of $r$ terms, and $\kappa_2=\frac{\xi^2}{r},\dots,\frac{\xi^2+r-1}{r},\frac{\alpha}{r},\dots,\frac{\alpha+r-1}{r}, \frac{m}{r},\dots,\frac{m+r-1}{r}$ comprises of $3r$ terms. The above CDF in \eqref{Eq:M_CDF} reduces to the CDF of Gamma-Gamma turbulence as
\begin{equation}\label{Eq:CDF_Single}
F_\gamma(\gamma)=J\,\MeijerG[right]{3r,1}{r+1,3r+1}{K\frac{\gamma}{\mu_{r}}}{1,\kappa_1}{\kappa_3,0},
\end{equation}
where $J=r^{\alpha+\beta-2}\,\xi^2/\left[(2\,\pi)^{r-1}\Gamma(\alpha)\Gamma(\beta)\right]$, $K=(\xi^{2}\alpha\,\beta)^r/\left[\left(\xi^{2}+1\right)^{r}r^{2\,r}\right]$, and $\kappa_3=\frac{\xi^2}{r},\dots,\frac{\xi^2+r-1}{r},\frac{\alpha}{r},\dots,\frac{\alpha+r-1}{r}, \frac{\beta}{r},\dots,\frac{\beta+r-1}{r}$ comprises of $3r$ terms. This unified expression for the CDF of a single unified FSO link in \eqref{Eq:CDF_Single} is in agreement (for $\xi^{2}\,>>1$) with the individual results presented in \cite[Eq. (15)]{nistazakis} (for $\xi \rightarrow \infty$ and $r=2$), \cite[Eq. (15)]{sandalidis} and \cite[Eq. (17)]{feng} (for $r=1$), \cite[Eq. (16)]{li-qiang} and \cite[Eq. (7)]{tsiftsis} (for $\xi \rightarrow \infty$ and $r=1$), and references cited therein. Mathematically, eq. \eqref{Eq:CDF_Single} can be easily derived from \eqref{Eq:M_CDF} by simply setting $\rho=1$ and $\Omega^{'}=1$ in \eqref{Eq:M_CDF}. Among all the sum terms in \eqref{Eq:M_CDF}, all terms become $0$ except for the term when $m=\beta$ \cite{samimi}. Hence, with this and with some simple algebraic manipulations, we can easily obtain \eqref{Eq:CDF_Single} from \eqref{Eq:M_CDF}.

\subsubsection{Asymptotic Analysis} Using \cite[Eq. (6.2.2)]{springer} to invert the argument in the Meijer's G function in \eqref{Eq:M_CDF} and then applying \eqref{Eq:G-Expansion} from the Appendix, the CDF for the $\mathcal{M}$ turbulence in \eqref{Eq:M_CDF} can be given asymptotically, at \textit{\textbf{high SNR}}, in a simpler form in terms of basic elementary functions as
\begin{equation}\label{Eq:MCDF}
\begin{aligned}
F_\gamma(\gamma)&\underset{\mu_{r}\,>>1}{\approxeq}D\sum_{m=1}^{\beta}c_{m}\sum_{k=1}^{3r}\left(\frac{\mu_{r}}{E\,\gamma}\right)^{-\kappa_{2,k}}\\
&\times\frac{\prod_{l=1;\,l\neq k}^{3r}\Gamma(\kappa_{2,l}-\kappa_{2,k})}{\kappa_{2,k}\prod_{l=2}^{r+1}\Gamma(\kappa_{1,l}-\kappa_{2,k})},
\end{aligned}
\end{equation}
where $\kappa_{u,v}$ represents the $v^{\mathrm{th}}$-term of $\kappa_{u}$. The asymptotic expression for the CDF in \eqref{Eq:MCDF} is dominated by the min$\left(\xi, \alpha, \beta\right)$ where $\xi$ represents the $1^{\mathrm{st}}$-term, $\alpha$ represents the $(r+1)^{\mathrm{th}}$-term, and $\beta$ represents the $(2\,r+1)^{\mathrm{th}}$-term in $\kappa_{2}$ i.e. when the difference between the parameters is greater than 1 then the asymptotic expression for the CDF in \eqref{Eq:MCDF} is dominated by a single term that has the least value among the above three parameters i.e. $\xi, \alpha$, and $\beta$. On the other hand, if the difference between any two parameters is less than 1 then the asymptotic expression for the CDF in \eqref{Eq:MCDF} is dominated by the summation of the two terms that have the least value among the above three parameters with a difference less than 1 and so on and so forth. As a special case, we can have the asymptotic CDF of the Gamma-Gamma turbulence as
\begin{equation}\label{Eq:CDF_Asymp}
\begin{aligned}
F_\gamma(\gamma)&\underset{\mu_{r}\,>>1}{\approxeq}J\sum_{k=1}^{3r}\left(\frac{\mu_{r}}{K\,\gamma}\right)^{-\kappa_{3,k}}\frac{\prod_{l=1;\,l\neq k}^{3r}\Gamma(\kappa_{3,l}-\kappa_{3,k})}{\kappa_{3,k}\prod_{l=2}^{r+1}\Gamma(\kappa_{1,l}-\kappa_{3,k})}.
\end{aligned}
\end{equation}

\subsection{Moment Generating Function}
\subsubsection{Exact Analysis} The MGF defined as $\mathcal{M}_{\gamma}(s) \triangleq \Expected{e^{-\gamma s}}$, can be expressed, using integration by parts, in terms of CDF as
\begin{equation}\label{Eq:MGFDef_CDF}
\mathcal{M}_{\gamma}(s) = s\int_{0}^{\infty}e^{-\gamma s} F_{\gamma}(\gamma)d\gamma.
\end{equation}
By placing \eqref{Eq:M_CDF} into \eqref{Eq:MGFDef_CDF} and utilizing \cite[Eq. (7.813.1)]{gradshteyn}, we get after some manipulations the MGF for the $\mathcal{M}$ turbulence as
\begin{equation}\label{Eq:M_MGF}
\mathcal{M}_{\gamma}(s)=D\sum_{m=1}^{\beta}c_{m}\,\MeijerG[right]{3r,2}{r+2,3r+1}{\frac{E}{\mu_{r}\,s}}{0,1,\kappa_{1}}{\kappa_{2},0}.
\end{equation}
As a special case, the MGF for the Gamma-Gamma turbulence is derived as
\begin{equation}\label{Eq:MGF_Single}
\mathcal{M}_{\gamma}(s)=J\,\MeijerG[right]{3r,2}{r+2,3r+1}{\frac{K}{\mu_{r}\,s}}{0,1,\kappa_1}{\kappa_3,0}.
\end{equation}
This unified expression for the MGF of a single unified FSO link in \eqref{Eq:MGF_Single} is in agreement (for $\xi^{2}\,>>1$) with the individual result presented in \cite[Eq. (3)]{peppas2} (for $\xi \rightarrow \infty$ and $r=2$), and references cited therein. Additionally, it is very important to note here that these results for the MGF are very much beneficial and handy as they can be utilized for MGF-based analysis and this can be seen from a very recent work in \cite{sharma} and references cited therein.

\subsubsection{Asymptotic Analysis} Similar to the CDF, the MGF for the $\mathcal{M}$ turbulence can be expressed asymptotically, at \textit{\textbf{high SNR}}, as
\begin{equation}\label{Eq:MMGF}
\begin{aligned}
\mathcal{M}_{\gamma}(s)&\underset{\mu_{r}\,>>1}{\approxeq}D\sum_{m=1}^{\beta}c_{m}\sum_{k=1}^{3r}\left(\frac{s}{E}\,\mu_{r}\right)^{-\kappa_{2,k}}\\
&\times\frac{\Gamma(\kappa_{2,k})\,\prod_{l=1;\,l\neq k}^{3r}\Gamma(\kappa_{2,l}-\kappa_{2,k})}{\prod_{l=3}^{r+2}\Gamma(\kappa_{1,l}-\kappa_{2,k})},
\end{aligned}
\end{equation}
for the Gamma-Gamma turbulence as
\begin{equation}\label{Eq:MGF_Asymp}
\begin{aligned}
\mathcal{M}_{\gamma}(s)&\underset{\mu_{r}\,>>1}{\approxeq}J\sum_{k=1}^{3r}\left(\frac{s}{K}\,\mu_{r}\right)^{-\kappa_{3,k}}\\
&\times\frac{\Gamma(\kappa_{3,k})\,\prod_{l=1;\,l\neq k}^{3r}\Gamma(\kappa_{3,l}-\kappa_{3,k})}{\prod_{l=3}^{r+2}\Gamma(\kappa_{1,l}-\kappa_{3,k})},
\end{aligned}
\end{equation}
and can be further expressed via only the dominant term(s) based on a similar explanation to the one given for the CDF case earlier.

\subsection{Moments}
The moments are defined as $\Expected{\gamma^n}$. Placing \eqref{Eq:M_PDF} into the definition and utilizing \cite[Eq. (7.811.4)]{gradshteyn}, we derive, to the best of our knowledge, a new expression for the moments of the $\mathcal{M}$ turbulence in exact closed-form and in terms of simple elementary functions as
\begin{equation}\label{Eq:M_Moments}
\Expected{\gamma^n}=\frac{r\,\xi^2\,A\,\Gamma(r\,n+\alpha)}{2^{r}\,\left(r\,n+\xi^2\right)\,B^{r\,n}}\sum_{m=1}^{\beta}b_{m}\,\Gamma(r\,n+m)\,\mu_{r}^{\,n},
\end{equation}
and of the Gamma-Gamma turbulence as
\begin{equation}\label{Eq:Moments_Single}
\Expected{\gamma^n}=\frac{\xi^2\left(\xi^{2}+1\right)^{r\,n}\Gamma(r\,n+\alpha)\Gamma(r\,n+\beta)}{\left(\xi^{2}\alpha\,\beta\right)^{r\,n}\left(r\,n+\xi^2\right)\Gamma(\alpha)\Gamma(\beta)}\,\mu_{r}^{\,n}.
\end{equation}
It is worthy to note that this simple result for the moments is particularly useful to conduct asymptotic analysis of the ergodic capacity in the later part of this work.

\section{Applications}
\subsection{Outage Probability}
When the instantaneous output SNR $\gamma$ falls below a given threshold $\gamma_{\mathrm{th}}$, we encounter a situation labeled as outage and it is an important feature to study OP of a system. Hence, another important fact worth stating here is that the expressions derived in \eqref{Eq:M_CDF} and \eqref{Eq:MCDF} also serve the purpose for the expressions of OP for a FSO channel or in other words, the probability that the SNR falls below a predetermined protection ratio $\gamma_{\mathrm{th}}$ can be simply expressed by replacing $\gamma$ with $\gamma_{\mathrm{th}}$ in \eqref{Eq:M_CDF} and \eqref{Eq:MCDF} as $P_{\rm{out}}(\gamma_{\rm{th}})=F_{\gamma}(\gamma_{\rm{th}})$. With $\gamma_{\rm{th}}=e^{2\,R}-1$, it must be noted here that the OP may be defined as the event that the capacity does not exceed the operating rate $R$ \cite{belmonte}.

\subsection{Scintillation Index}
The SI is an important measure for the performance of optical wireless communication systems. In particular, the SI is defined as $\sigma_{I}^{2}=\mathbb{E}_{I}[I^{2}]/\mathbb{E}_{I}^{2}[I]-1$ \cite{andrews}, \cite[Eq. (6)]{niu1}.
Now, substituting \eqref{Eq:M_Moments} appropriately into this definition, we can easily get the exact closed-form expression for the SI.

\subsection{Average BER}
\subsubsection{Exact Analysis} Substituting \eqref{Eq:M_CDF} into \cite[Eq. (12)]{ansari} and utilizing \cite[Eq. (7.813.1)]{gradshteyn}, we get the average BER $\overline{P}_b$ of a variety of binary modulations for the $\mathcal{M}$ turbulence as
\begin{equation}\label{Eq:M_BER}
\overline{P}_b=\frac{D}{2\,\Gamma(p)}\sum_{m=1}^{\beta}c_{m}\,\MeijerG[right]{3r,2}{r+2,3r+1}{\frac{E}{\mu_{r}\,q}}{1-p,1,\kappa_1}{\kappa_2,0},
\end{equation}
and for the Gamma-Gamma turbulence as
\begin{equation}\label{Eq:BER_Single}
\overline{P}_b=\frac{J}{2\,\Gamma(p)}\,\MeijerG[right]{3r,2}{r+2,3r+1}{\frac{K}{\mu_{r}\,q}}{1-p,1,\kappa_1}{\kappa_3,0},
\end{equation}
where the parameters $p$ and $q$ account for different modulation schemes. For an extensive list of modulation schemes represented by these parameters, one may look into \cite{sagias1,wojnar,ansari,ansari4} or refer to \tableref{Table:ModulationSchemes}.
\begin{table}[t]
\begin{center}
\caption{BER Parameters of Binary Modulations}
\label{Table:ModulationSchemes}
\begin{tabular}{l c c}\hline\hline\\[-1mm]
\textbf{Modulation} & \textbf{$p$} & \textbf{$q$} \\[1mm]\hline\\[0mm]
Coherent Binary Frequency Shift Keying (CBFSK) & 0.5 & 0.5\\[1mm]
Coherent Binary Phase Shift Keying (CBPSK) & 0.5 & 1\\[1mm]
Non-Coherent Binary Frequency Shift Keying (NBFSK) & 1 & 0.5\\[1mm]
Differential Binary Phase Shift Keying (DBPSK) & 1 & 1\\[1mm]\hline
\end{tabular}
\end{center}
\end{table}
This unified expression for the BER of a single unified FSO link in \eqref{Eq:BER_Single} is in agreement (for $\xi^{2}\,>>1$) with the individual results presented in \cite[Eq. (5)]{song} (for $r=2$), \cite[Eq. (24)]{sandalidis} (for $r=1$), \cite[Eq. (10)]{tsiftsis} and \cite[Eq. (7)]{niu} (for $\xi \rightarrow \infty$ and $r=1$), and references cited therein.

\subsubsection{Asymptotic Analysis} Similar to the CDF, the BER can be expressed asymptotically for the $\mathcal{M}$ turbulence, at \textit{\textbf{high SNR}}, as
\begin{equation}\label{Eq:MBER}
\begin{aligned}
\overline{P}_b&\underset{\mu_{r}\,>>1}{\approxeq}\frac{D}{2\,\Gamma(p)}\sum_{m=1}^{\beta}c_{m}\sum_{k=1}^{3r}\left(\frac{q}{E}\,\mu_{r}\right)^{-\kappa_{2,k}}\\
&\times\frac{\Gamma(\kappa_{2,k}+p)\,\prod_{l=1;\,l\neq k}^{3r}\Gamma(\kappa_{2,l}-\kappa_{2,k})}{\kappa_{2,k}\,\prod_{l=3}^{r+2}\Gamma(\kappa_{1,l}-\kappa_{2,k})},
\end{aligned}
\end{equation}
for the Gamma-Gamma turbulence as
\begin{equation}\label{Eq:BER_Asymp}
\begin{aligned}
\overline{P}_b&\underset{\mu_{r}\,>>1}{\approxeq}\frac{J}{2\,\Gamma(p)}\sum_{k=1}^{3r}\left(\frac{q}{K}\,\mu_{r}\right)^{-\kappa_{3,k}}\\
&\times\frac{\Gamma(\kappa_{3,k}+p)\,\prod_{l=1;\,l\neq k}^{3r}\Gamma(\kappa_{3,l}-\kappa_{3,k})}{\kappa_{3,k}\,\prod_{l=3}^{r+2}\Gamma(\kappa_{1,l}-\kappa_{3,k})},
\end{aligned}
\end{equation}
and can be further expressed via only the dominant term(s) based on a similar explanation to the one given for the CDF case earlier.
\subsubsection{Diversity Order and Coding Gain} Utilizing $\overline{P}_b\approx \left(G_{c}\,\mu_{r}\right)^{-G_{d}}$ \cite[Eq. (1)]{wang}, we can easily share for the $\mathcal{M}$ turbulence, the diversity order is $G_{d}=$ min$\left(\xi^{2}/r, \alpha/r, \beta/r\right)\underset{\alpha>\beta}{=}$ min$\left(\xi^{2}/r, \beta/r\right)$ and the coding gain is
\begin{equation}\label{Eq:MGc}
\begin{aligned}
G_{c}&=q/E\left(D/\left(2\,\Gamma(p)\right)\sum_{m=1}^{\beta}c_{m}\right.\\
&\times\left.\frac{\Gamma(\kappa_{2,k}+p)\,\prod_{l=1;\,l\neq k}^{3r}\Gamma(\kappa_{2,l}-\kappa_{2,k})}{\kappa_{2,k}\,\prod_{l=3}^{r+2}\Gamma(\kappa_{1,l}-\kappa_{2,k})}\right)^{-\frac{1}{\kappa_{2,k}}}.
\end{aligned}
\end{equation}
Similarly, for the Gamma-Gamma turbulence, we have diversity order as $G_{d}=$ min$\left(\xi^{2}/r, \alpha/r, \beta/r\right)\underset{\alpha>\beta}{=}$ min$\left(\xi^{2}/r, \beta/r\right)$ and the coding gain as
\begin{equation}\label{Eq:Gc}
\begin{aligned}
G_{c}&=q/K\left(J/\left(2\,\Gamma(p)\right)\right.\\
&\times\left.\frac{\Gamma(\kappa_{3,k}+p)\,\prod_{l=1;\,l\neq k}^{3r}\Gamma(\kappa_{3,l}-\kappa_{3,k})}{\kappa_{3,k}\,\prod_{l=3}^{r+2}\Gamma(\kappa_{1,l}-\kappa_{3,k})}\right)^{-\frac{1}{\kappa_{3,k}}}.
\end{aligned}
\end{equation}
The authors in \cite[Eqs. (18) and (19)]{bayaki} have also derived the diversity order and the coding gain for the Gamma-Gamma turbulent FSO channels though applicable to the IM/DD technique under negligible pointing errors. Hence, the results for the coding gain and the diversity order in this work are applicable to even the heterodyne detection technique and also for non-negligible pointing error effects for both types of detection techniques.

\subsection{Average SER}
In \cite{alouini2}, the conditional SER has been presented in a desirable form and utilized to obtain the average SER of M-AM, M-PSK, and M-QAM. For example, for M-PSK the average SER $\overline{P}_s$ over generalized fading channels is given by \cite[Eq. (41)]{alouini2}. Similarly, for M-AM and M-QAM, the average SER $\overline{P}_s$ over generalized fading channels is given by \cite[Eq. (45)]{alouini2} and \cite[Eq. (48)]{alouini2} respectively. On substituting \eqref{Eq:M_MGF} into \cite[Eq. (41)]{alouini2}, \cite[Eq. (45)]{alouini2}, and \cite[Eq. (48)]{alouini2}, we can get the SER of M-PSK, M-AM, and M-QAM, respectively. The analytical SER performance expressions obtained via the above substitutions are exact and can be easily estimated accurately by utilizing the Gauss-Chebyshev Quadrature (GCQ) formula \cite[Eq. (25.4.39)]{abramowitz} that converges rapidly, requiring only few terms for an accurate result \cite{yilmaz2}.

\subsection{Ergodic Capacity}
It is well known that the atmospheric turbulence over FSO links is slow in fading. Since the coherence time of the channel is in the order of milliseconds (ms), turbulence induced fading remains constant over a large number of transmitted bits \cite{lee1,farid,chatzidiamantis1,petkovic}. In addition, the use of very long inter-leavers in order to achieve independent fading samples in consecutive symbol intervals is not practical in FSO channel \cite{farid}. Moreover, in this study we are including the effects of the pointing error that makes the signal fluctuate at a very high rate. Because the coherence time of the FSO fading channel is in the order of milliseconds, a single fade can obliterate millions of bits at Gbits/s (Gbps) data rates and therefore, the average (i.e. ergodic) capacity of the channel, represents the best achievable capacity of an optical wireless link. Hence, our aim is to examine the best possible ergodic performance. Therefore, our ergodic capacity analysis is valid under the presence of the pointing errors and under the assumption that the information symbol is long enough to ensure the long-term ergodic properties of the turbulence process \cite{chatzidiamantis1,petkovic}.
\subsubsection{Exact Analysis} The ergodic channel capacity $\overline{C}$ is defined as $\overline{C} \triangleq \Expected{\log_2(1+c\,\gamma)}$ where $c$ is a constant term such that $c=1$ for heterodyne detection and $c=e/\left(2\,\pi\right)$ for IM/DD \cite[Eq. (26)]{lapidoth},\cite[Eq. (7.43)]{owc}. Utilizing this equation by placing \eqref{Eq:MPDF} in it, using \cite[Eq. (07.34.03.0456.01)]{mathematica} to represent $\ln(1+c\,\gamma)$ in terms of Meijer's G function as $\MeijerG[right]{1,2}{2,2}{c\,\gamma}{1,1}{1,0}$, and using \cite[Eq. (21)]{adamchik}, the ergodic capacity for the $\mathcal{M}$ turbulence can be expressed as
\begin{equation}\label{Eq:M_EC}
\overline{C}=\frac{D}{\ln(2)}\sum_{m=1}^{\beta}c_{m}\,\MeijerG[right]{3r+2,1}{r+2,3r+2}{\frac{E}{\mu_{r}^{\sim}}}{0,1,\kappa_1}{\kappa_2,0,0},
\end{equation}
and for the Gamma-Gamma turbulence as
\begin{equation}\label{Eq:EC_Single}
\overline{C}=\frac{J}{\ln(2)}\,\MeijerG[right]{3r+2,1}{r+2,3r+2}{\frac{K}{\mu_{r}^{\sim}}}{0,1,\kappa_1}{\kappa_3,0,0},
\end{equation}
where $\mu_{r}^{\sim}=c\,\mu_{r}$. Specifically, $\mu_{1}^{\sim}=\mu_{1}$ for the heterodyne detection technique (i.e. $r=1$) and $\mu_{2}^{\sim}=e/\left(2\,\pi\right)\,\mu_{2}$ for the IM/DD technique (i.e. $r=2$). The unified expression for the ergodic capacity of a single unified $\mathcal{M}$ FSO link in \eqref{Eq:M_EC} is in agreement (for $r=2$) with the individual result presented in \cite[Eq. (8)]{balsells1}. Also, the unified expression for the ergodic capacity of a single unified Gamma-Gamma FSO link in \eqref{Eq:EC_Single} is in agreement (for $\xi^{2}\,>>1$) with the individual results presented in \cite[Eq. (22)]{gappmair} (for $r=2$), \cite[Eq. (21)]{nistazakis} and \cite[Eq. (11)]{gappmair} (for $\xi \rightarrow \infty$ and $r=2$), \cite[Eq. (10)]{liu2} (for $r=1$), \cite[Eq. (16)]{nistazakis2} and \cite[Eq. (3)]{ren} (for $\xi \rightarrow \infty$ and $r=1$), and references cited therein.

For readers clarification, the Shannon ergodic capacity given as $\overline{C} \triangleq \Expected{\log_2(1+c\,\gamma)}$ is true as an exact expression for deriving the respective ergodic capacity for heterodyne detection technique whereas for IM/DD technique, it acts as a lower bound as given in \cite[Eq. (26)]{lapidoth} and \cite[Eq. (7.43)]{owc}. Hence, we can safely claim that the ergodic capacity's derived in \eqref{Eq:M_EC} and \eqref{Eq:EC_Single} above are an exact solution for the heterodyne detection technique (i.e. $r=1$) whereas for the IM/DD technique (i.e. $r=2$), these solutions (as per above mentioned references) act as a lower bound.

\subsubsection{Asymptotic Analysis} Similar to the CDF, the ergodic capacity for the $\mathcal{M}$ turbulence can be expressed asymptotically via utilizing the Meijer's G function expansion given in the Appendix, at \textit{\textbf{high SNR}}, as
\begin{equation}\label{Eq:MCAP}
\begin{aligned}
\overline{C}&\underset{\mu_{r}\,>>1}{\approxeq}\frac{D}{\ln(2)}\sum_{m=1}^{\beta}c_{m}\sum_{k=1}^{3r+2}\left(\frac{\mu_{r}^{\sim}}{E}\right)^{-\kappa_{2,k}}\\
&\times\frac{\Gamma(1+\kappa_{2,k})\prod_{l=1;\,l\neq k}^{3r+2}\Gamma(\kappa_{2,l}-\kappa_{2,k})}{\Gamma(1-\kappa_{2,k})\prod_{l=3}^{r+2}\Gamma(\kappa_{1,l}-\kappa_{2,k})},
\end{aligned}
\end{equation}
for the Gamma-Gamma turbulence as
\begin{equation}\label{Eq:CAP_Asymp}
\begin{aligned}
\overline{C}&\underset{\mu_{r}\,>>1}{\approxeq}\frac{J}{\ln(2)}\sum_{k=1}^{3r+2}\left(\frac{\mu_{r}^{\sim}}{K}\right)^{-\kappa_{3,k}}\\
&\times\frac{\Gamma(1+\kappa_{3,k})\prod_{l=1;\,l\neq k}^{3r+2}\Gamma(\kappa_{3,l}-\kappa_{3,k})}{\Gamma(1-\kappa_{3,k})\prod_{l=3}^{r+2}\Gamma(\kappa_{1,l}-\kappa_{3,k})},
\end{aligned}
\end{equation}
and can be further expressed via only the dominant term(s) based on the similar explanation as given for the CDF case earlier except with min$\left(\xi, \alpha, \beta,1,1+\epsilon\right)$ instead of min$\left(\xi, \alpha, \beta\right)$, where $\epsilon$ is a very small error introduced so as not to violate the conditions given in the Appendix, required to utilize \eqref{Eq:G-Expansion}.

Alternatively, a \textit{\textbf{high SNR}} asymptotic analysis may also be done by utilizing the moments as \cite[Eqs. (8) and (9)]{yilmaz3}
\begin{equation}\label{Eq:EC_Single_High_Def}
\overline{C}\underset{\mu_{r}\,>>1}{\approxeq}\log(\mu_{r})+\zeta,
\end{equation}
where
\begin{equation}\label{Eq:AF_Single_Derivative_Def}
\zeta=\left.\frac{\partial}{\partial n}AF_\gamma^{(n)}\right|_{n=0}.
\end{equation}
The expression in \eqref{Eq:EC_Single_High_Def} can be simplified to
\begin{equation}\label{Eq:EC_Single_High_Def_Simplify}\small
\overline{C}\underset{\mu_{r}\,>>1}{\approxeq}\ln(c\,\mu_{r})+\left.\frac{\partial}{\partial n}\left(\frac{\Expected{\gamma_{r}^n}}{\Expected{\gamma_{r}}^n}-1\right)\right|_{n=0}=\left.\frac{\partial}{\partial n}\,\Expected{\gamma_{r}^n}\right|_{n=0}.
\end{equation}\normalsize
Hence, we need to evaluate the first derivative of the moments in \eqref{Eq:M_Moments} at $n=0$ for high SNR asymptotic approximation to the ergodic capacity in \eqref{Eq:M_EC}. The first derivative of the moments is given as
\begin{equation}\label{Eq:M_Moments_Derivative}
\begin{aligned}
\frac{\partial}{\partial n}\,\Expected{\gamma^n}&=\frac{r\,\xi^2\,A\,\Gamma(r\,n+\alpha)}{2^{r}\,\left(r\,n+\xi^2\right)\,B^{r\,n}}\sum_{m=1}^{\beta}b_{m}\,\Gamma(r\,n+m)\\
&\times\left\{r\left[\psi(r\,n+\alpha)+\,\psi(r\,n+m)-\log(B)\right]\right.\\
&\left.+\log(\mu_{r})-r/\left(r\,n+\xi^2\right)\right\}\,\mu_{r}^{\sim\,n},
\end{aligned}
\end{equation}
where $\psi(.)$ is the digamma (psi) function \cite[Eq. (6.3.1)]{abramowitz}, \cite[Eq. (8.360.1)]{gradshteyn}. Evaluating \eqref{Eq:M_Moments_Derivative} at $n=0$, we obtain
\begin{equation}\label{Eq:M_Moments_Derivative_n0}
\begin{aligned}
\overline{C}&\underset{\mu_{r}\,>>1}{\approxeq}\frac{r\,A\,\Gamma(\alpha)}{2^{r}}\sum_{m=1}^{\beta}b_{m}\,\Gamma(m)\\
&\times\left\{r\left[-1/\xi^2-\log(B)+\psi(\alpha)+\,\psi(m)\right]+\log(\mu_{r}^{\sim})\right\}.
\end{aligned}
\end{equation}
Hence, eq. \eqref{Eq:M_Moments_Derivative_n0} gives the required expression for $\overline{C}$ for the $\mathcal{M}$ turbulent channel at high SNR in terms of simple elementary functions. Similar expression is derived for the Gamma-Gamma turbulent channel as
\begin{equation}\label{Eq:Moments_Derivative_n0}
\begin{aligned}
\overline{C}\underset{\mu_{r}\,>>1}{\approxeq}&\log(\mu_{r}^{\sim})+\,r\left[-\frac{1}{\xi^2}-\log\left(\frac{\xi^2}{\xi^2+1}\right)\right.\\
&\left.-\log(\alpha\,\beta)+\psi(\alpha)+\,\psi(\beta)\right].
\end{aligned}
\end{equation}

Furthermore, for \textit{\textbf{low SNR}} asymptotic analysis, it can be easily shown that the ergodic capacity can be asymptotically approximated by the first moment. We can utilize \eqref{Eq:M_Moments} via placing $n=1$ in it and hence the ergodic capacity of a single FSO link under $\mathcal{M}$ turbulence can be approximated at low SNR in closed-form in terms of simple elementary functions as
\begin{equation}\label{Eq:M_EC_Single_Low}
\overline{C}\underset{\mu_{r}\,<<1}{\approxeq}\Expected{\gamma^{n=1}}=\frac{r\,\xi^2\,A\,\Gamma(r+\alpha)}{2^{r}\,\left(r+\xi^2\right)\,B^{r}}\sum_{m=1}^{\beta}b_{m}\,\Gamma(r+m)\,\mu_{r}^{\sim},
\end{equation}
and under Gamma-Gamma turbulence as
\begin{equation}\label{Eq:EC_Single_Low}
\overline{C}\underset{\mu_{r}\,<<1}{\approxeq}\Expected{\gamma^{n=1}}=\frac{\xi^2\left(\xi^{2}+1\right)^{r}\,\Gamma(r+\alpha)\Gamma(r+\beta)}{\left(\xi^{2}\alpha\,\beta\right)^{r}\left(r+\xi^2\right)\Gamma(\alpha)\Gamma(\beta)}\,\mu_{r}^{\sim}.
\end{equation}

\section{Numerical Results and Discussion}
The FSO link is modeled as $\mathcal{M}$ turbulent channel with a link length of $L=1$ km and wavelength of $\lambda=785$ nm that contributes in obtaining $k_{w}=2\,\pi/\lambda$. The refraction structure parameter $C_{n}^{2}$ is considered as $C_{n}^{2}=1.2\times 10^{-13}\,\rm{m}^{-2/3}$, $C_{n}^{2}=10^{-11}\,\rm{m}^{-2/3}$, and $C_{n}^{2}=2.8\times 10^{-14}\,\rm{m}^{-2/3}$. These are then utilized to obtain the Rytov variance $\sigma_{R}^{2}=1.23\,C_{n}^{2}\,k_{w}^{7/6}\,L^{11/6}$ that subsequently define the effects of atmosphere as ($\alpha=2.296$; $\beta=2$), ($\alpha=4.2$; $\beta=3$) and ($\alpha=8$; $\beta=4$), ($\Omega=1.3265$, $b_{0}=0.1079$), $\rho=0.596$, and $\phi_{A}-\phi_{B}=\pi/2$. \footnote{It is important to note here that these values for the parameters were selected from \cite{navas,navas1,samimi} subject to the standards to prove the validity of the obtained results and hence other specific values can be used to obtain the required results by design communication engineers before deployment. Also, for all cases, $10^{6}$ realizations of the random variable were generated to perform the Monte-Carlo simulations in MATLAB.} In MATLAB, a $\mathcal{M}$ turbulent channel random variable was generated via squaring the absolute value of a Rician-shadowed random variable \cite{navas}.

\begin{figure}[h]
\begin{center}
\includegraphics[scale=0.29]{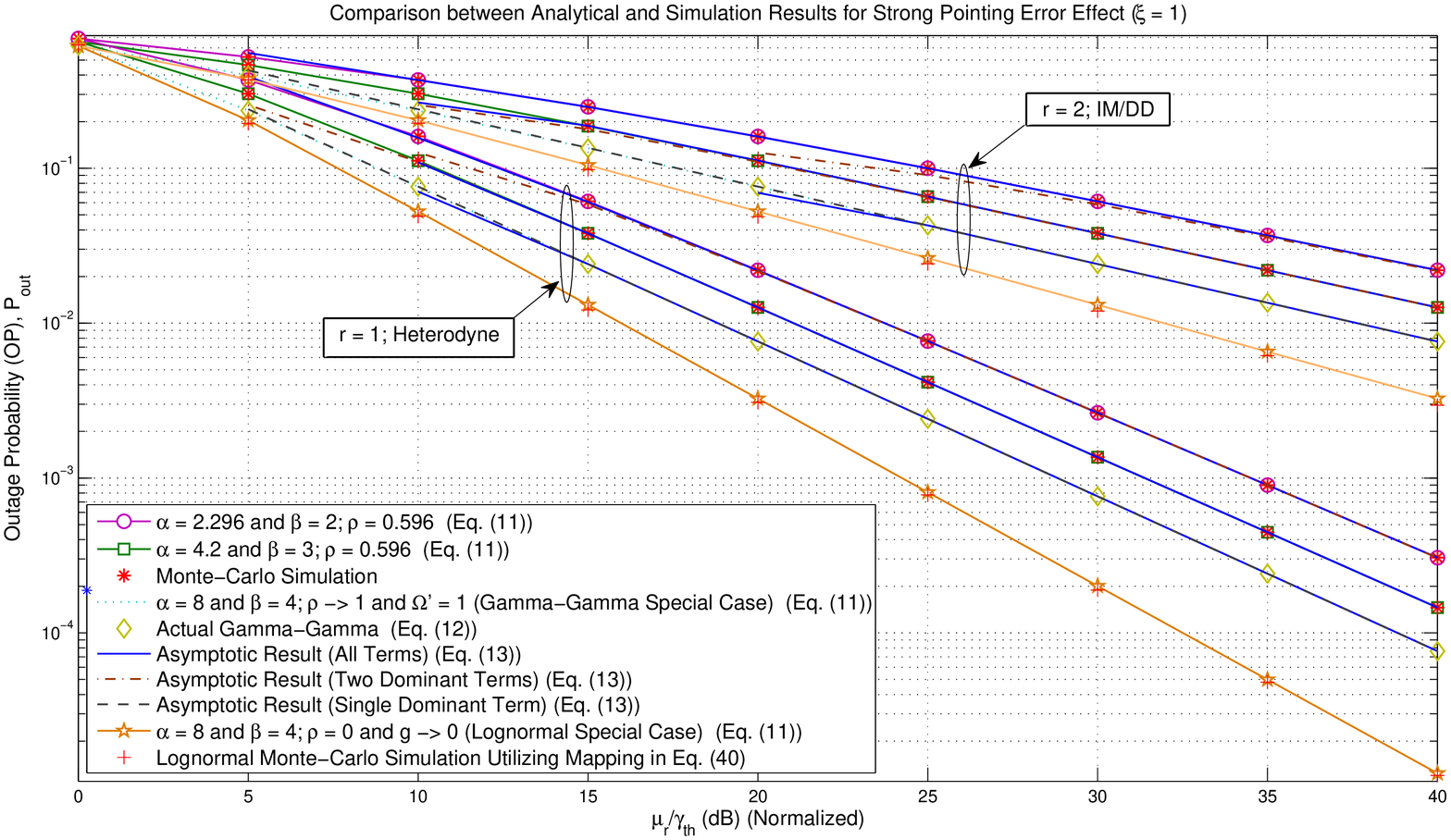}
\caption{OP showing the performance of both the detection techniques (heterodyne and IM/DD) under different turbulence conditions.}
\label{Fig:M_OP_r12}
\end{center}
\end{figure}

The OP is presented in \figref{Fig:M_OP_r12} for both types of detection techniques (i.e. IM/DD and heterodyne) across the normalized electrical SNR with fixed effect of the pointing error ($\xi = 1$).
We can observe from \figref{Fig:M_OP_r12} that the simulation results provide a perfect match to the analytical results obtained in this work. Additionally, it can be observed that as the effect of atmospheric turbulence decreases, the performance improves. It can be seen that at high SNR, the asymptotic expression derived in \eqref{Eq:MCDF} (i.e. utilizing all the terms in the summation) converges quite fast to the exact result proving this asymptotic approximation to be tight enough. Based on the effects of the turbulent parameters and the pointing error, the appropriate dominant term(s) can be selected as has been discussed earlier under the CDF subsection. Hence, we can see that these respective dominant term(s) also converge though relatively slower, specially for the IM/DD technique. More importantly, we can observe that once we apply $\rho=1$ and $\Omega^{'}=1$, the $\mathcal{M}$ turbulence matches exactly the special case of Gamma-Gamma turbulence. This can be depicted from the case wherein ($\alpha=8$; $\beta=4$).

Furthermore, on applying $\rho=0$ and $g\rightarrow 0$ to $\mathcal{M}$ atmospheric turbulence, one can obtain an approximation to weak lognormal atmospheric turbulence \cite{navas}. Hence, to analyze this, we derived the mapping for the lognormal parameter $\sigma_{I}$ in terms of the parameters of $\mathcal{M}$ atmospheric turbulence i.e. in terms of $\alpha, \beta, \xi, m, \Omega^{'},$ and $g$. Specifically, $\sigma_{I}$ was obtained via the moment matching method. The moments of lognormal turbulence are given as $\Expected{I^n}_{LN}=\frac{\xi^{2(1-r\,n)}}{\left(\xi^{2}+r\,n\right)\left(\xi^{2}+1\right)^{-r\,n}}\,\exp\left\{\frac{r\,n\,\sigma_{I}^{2}}{2}\left(r\,n-1\right)\right\}$ \cite[Eq. (18)]{ansari12} and the moments for the $\mathcal{M}$ turbulence can be easily extracted from \eqref{Eq:M_Moments}. On matching the second moment, we obtained the mapping for $\sigma_{I}$ as
\begin{equation}\label{Eq:sigma}\small
\sigma_{I}=\sqrt{\frac{1}{r\,\left(2\,r-1\right)}\ln\left\{\frac{r\,\xi^{4\,r}\,A\,\Gamma\left(\alpha+2\,r\right)}{2^{\,r}\,\left(\xi^{2}+1\right)^{2\,r}\,B^{2\,r}}\sum_{m=1}^{\beta}b_{m}\,\Gamma\left(m+2\,r\right)\right\}}.
\end{equation}\normalsize
The plot for this scenario can be easily depicted in \figref{Fig:M_OP_r12} from the case wherein ($\alpha=8$; $\beta=4$). It must be noted that the curve signified by the second last entry in the legend depicts the lognormal special case approximate plotted via utilizing the unified exact closed-form OP analytical expression in \eqref{Eq:M_CDF}. The last entry in the legend of \figref{Fig:M_OP_r12} depicts the Monte-Carlo simulation/generation for lognormal random variable with $\sigma_{I}$ acquiring values from \eqref{Eq:sigma}. The values for the lognormal variance were obtained as $\sigma_{I}^{2}=0.3409$ for $r=1$ (i.e. for heterodyne FSO systems) scenario and as $\sigma_{I}^{2}=0.3079$ for $r=2$ scenario (i.e. for IM/DD FSO systems), respectively. It can be clearly observed that this approximation of $\mathcal{M}$ turbulence to lognormal is quite tight. Moreover, we had obtained expressions for $\sigma_{I}$ via matching the first moment as well and it was realized that the expression derived via matching the second moment gave tighter approximate results. Based on this, we can easily conclude that the higher moments we utilize to derive the mapping expression for $\sigma_{I}$, the tighter approximate may be obtained.

Additionally, another important outcome must be observed that the heterodyne detection technique, being more complex method of detection technique, performs better than the IM/DD technique. For instance, for $\alpha=2.296$, $\beta=2$, and $\rho=0.596$, at an electrical SNR of $15$ dB, the heterodyne detection technique outperforms the IM/DD technique in terms of the OP by $1.8852 * 10^{-1}$. On the other hand, for $\alpha=8$, $\beta=4$, $\rho\rightarrow 1$, and $\Omega^{'}=1$, for a desired OP i.e. lets say for $P_{\rm{out}} = 7.6 * 10^{-3}$, the heterodyne detection technique outperforms the IM/DD technique by $20$ dB.

\begin{figure}[h]
\begin{center}
\includegraphics[scale=0.29]{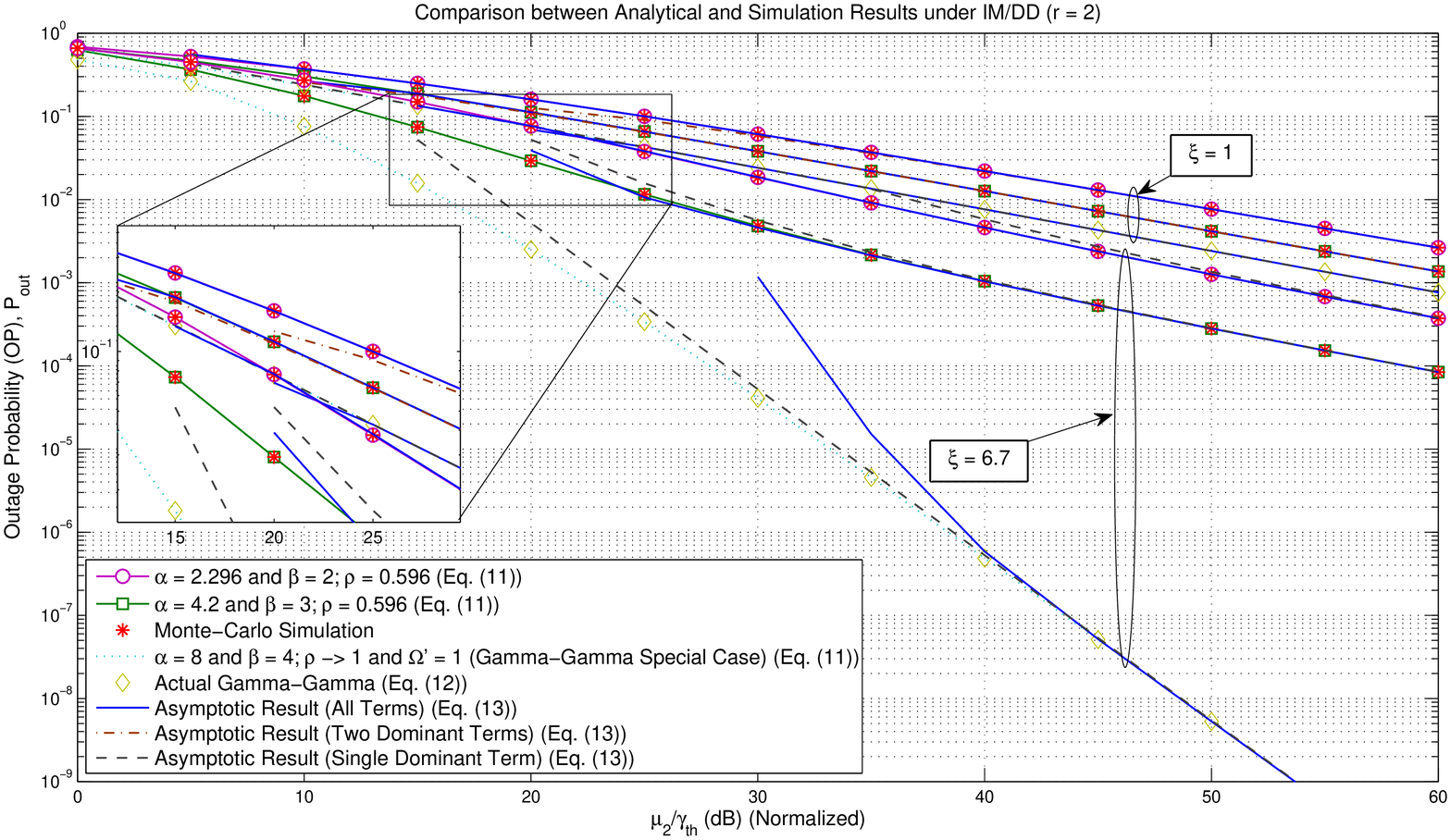}
\caption{OP showing the performance of IM/DD technique under different turbulence conditions with varying effects of pointing error.}
\label{Fig:M_OP_r2_xi}
\end{center}
\end{figure}

Similarly, \figref{Fig:M_OP_r2_xi} presents the OP for varying effects of pointing error ($\xi = 1~\mathrm{and}~6.7$) under the IM/DD technique.
We can observe that for lower effect of the pointing error (i.e. higher value of $\xi$), the respective performance gets better manifolds. Other outcomes, specially for the asymptotic approximations, can be observed similar to \figref{Fig:M_OP_r12} above except when the atmospheric effects get weaker and weaker wherein the single dominant term of  the asymptotic result converges faster than the sum of all terms in the asymptotic result.

\begin{figure}[h]
\begin{center}
\includegraphics[scale=0.29]{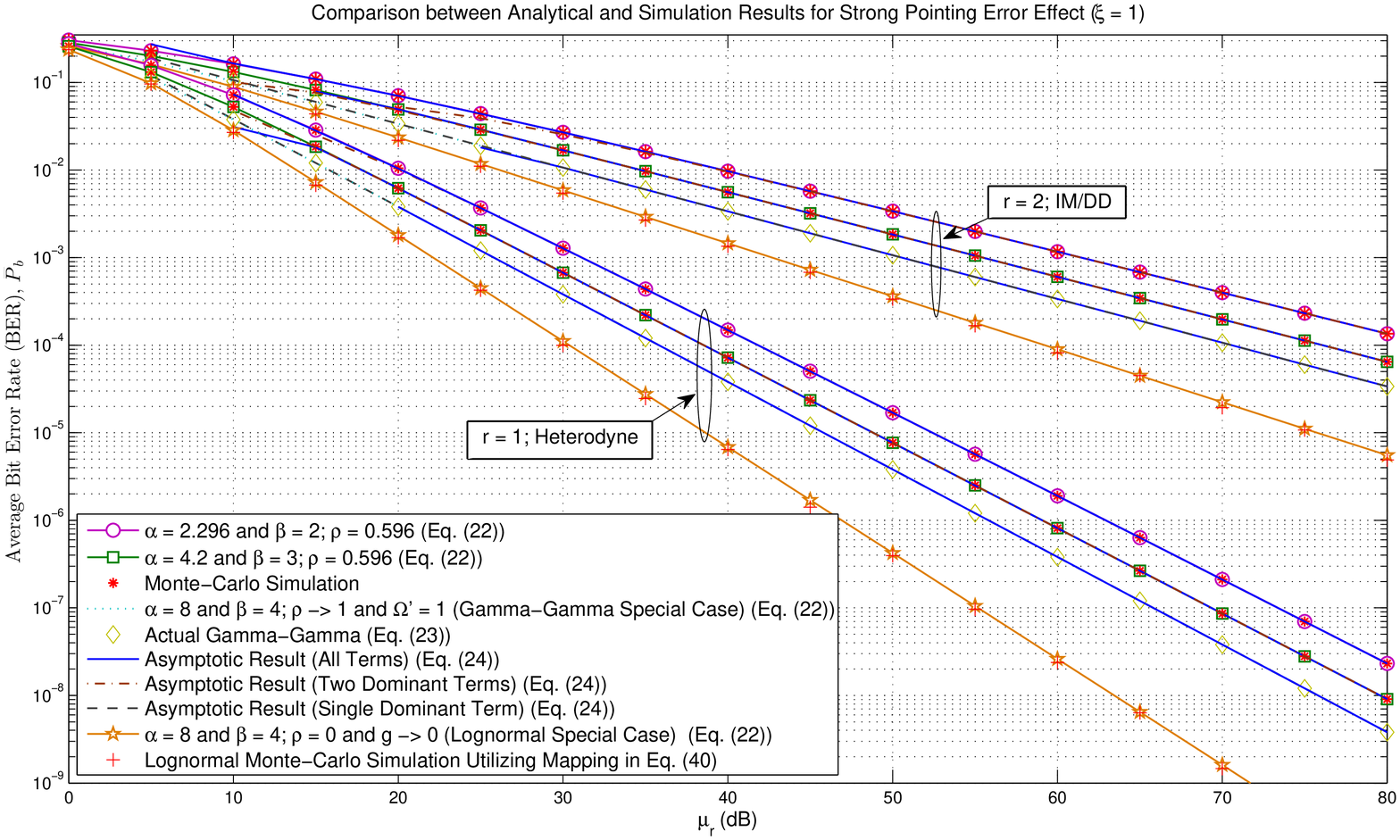}
\caption{Average BER of DBPSK binary modulation scheme showing the performance of both the detection techniques (heterodyne and IM/DD) under different turbulence conditions.}
\label{Fig:M_BER_r12}
\end{center}
\end{figure}

\begin{figure}[h]
\begin{center}
\includegraphics[scale=0.29]{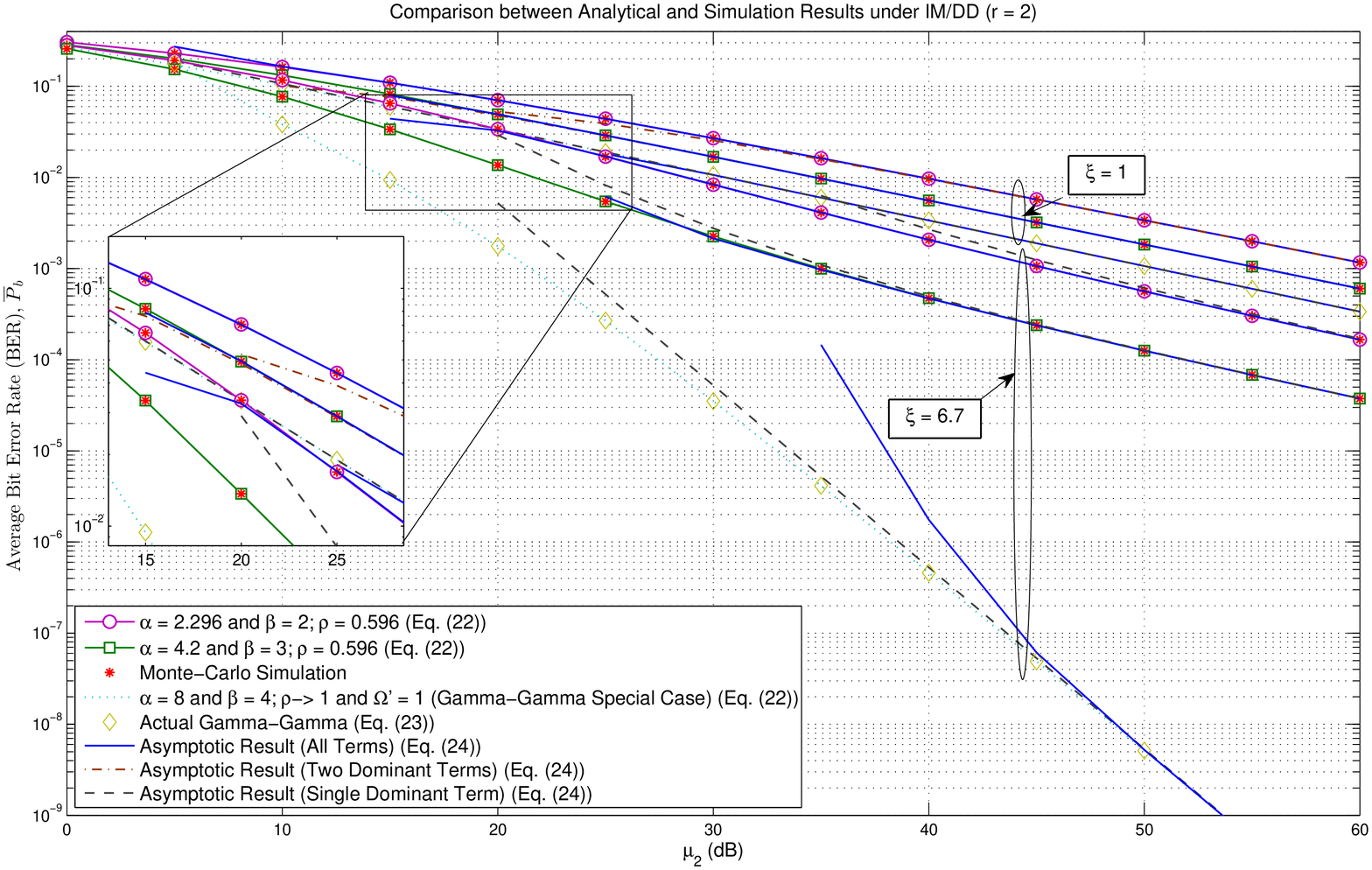}
\caption{Average BER of DBPSK binary modulation scheme showing the performance of IM/DD technique under different turbulence conditions for varying effects of pointing error.}
\label{Fig:M_BER_r2_xi}
\end{center}
\end{figure}

The average BER performance of DBPSK binary modulation scheme is presented in \figref{Fig:M_BER_r12}.
The effect of pointing error is fixed at $\xi=1$. Similar results can be observed as were observed for \figref{Fig:M_OP_r12}. Similarly, \figref{Fig:M_BER_r2_xi} presents the average BER for varying effects of pointing error ($\xi = 1~\mathrm{and}~6.7$) under the IM/DD technique.
We can observe that for lower effect of the pointing error ($\xi \rightarrow \infty$), the respective performance gets better. Other outcomes, specially for the asymptotic approximations, can be observed similar to \figref{Fig:M_OP_r2_xi} above.

\begin{figure}[h]
\begin{center}
\includegraphics[scale=0.29]{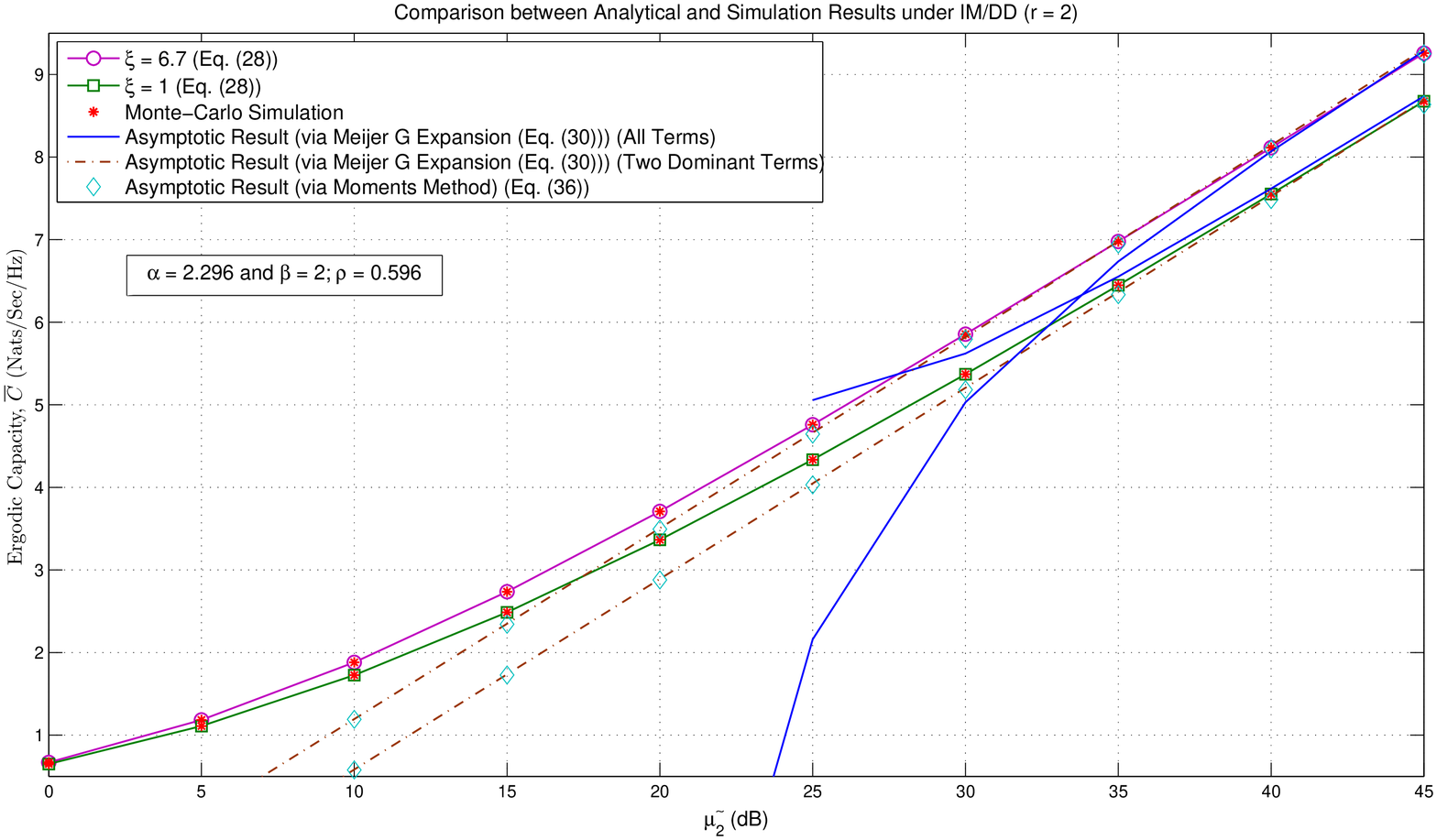}
\caption{Ergodic capacity results for the IM/DD technique with varying pointing errors along with the asymptotic results in high SNR regime for strong atmospheric turbulence.}
\label{Fig:M_EC_r2_S}
\end{center}
\end{figure}

\begin{figure}[h]
\begin{center}
\includegraphics[scale=0.29]{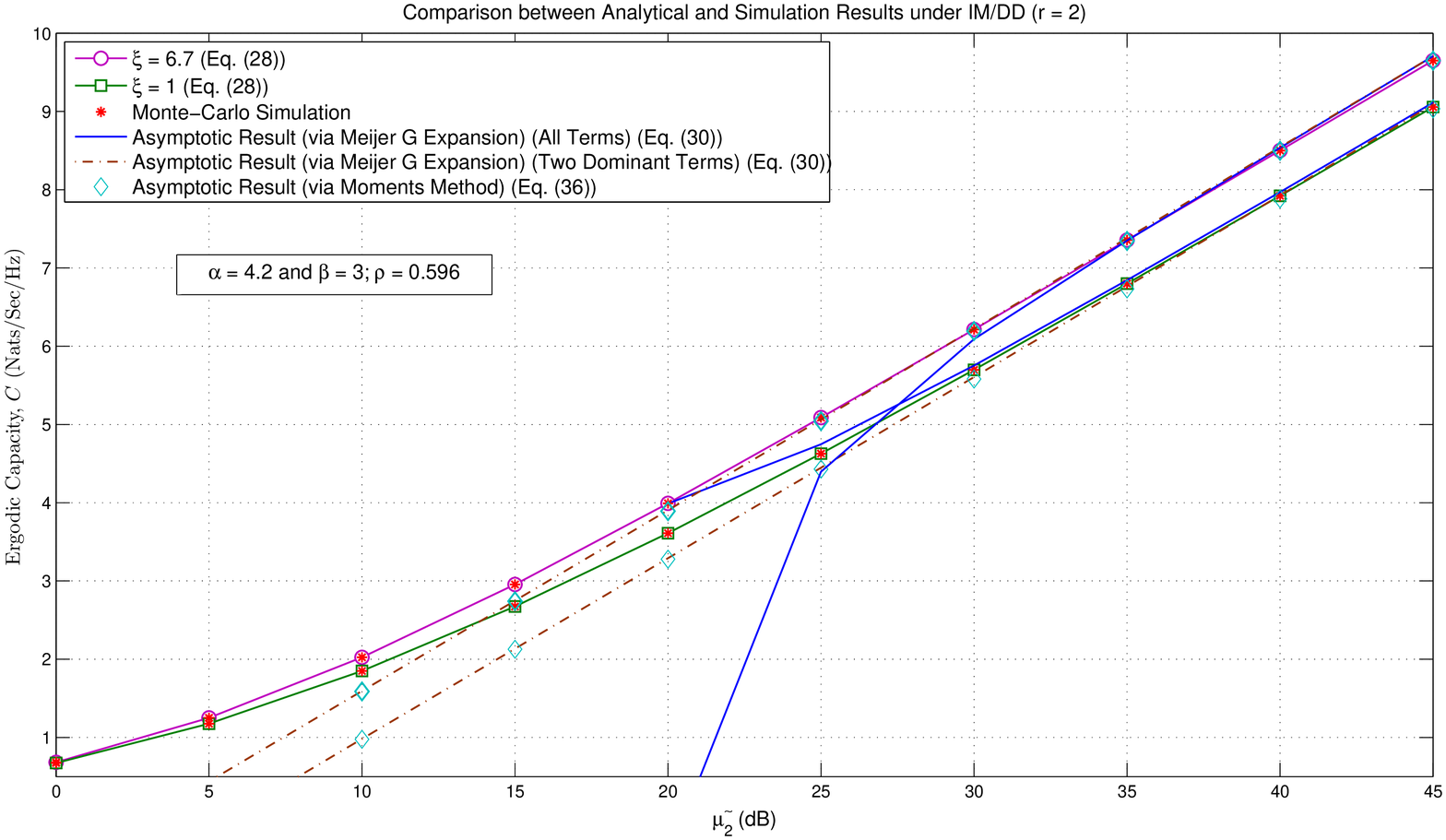}
\caption{Ergodic capacity results for the IM/DD technique with varying pointing errors along with the asymptotic results in high SNR regime for weak atmospheric turbulence.}
\label{Fig:M_EC_r2_W}
\end{center}
\end{figure}

\begin{figure}[h]
\begin{center}
\includegraphics[scale=0.29]{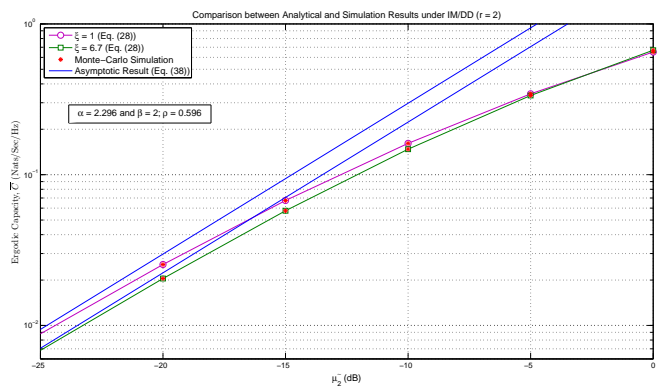}
\caption{Ergodic capacity results for the IM/DD technique for varying pointing errors along with the asymptotic results in low SNR regime.}
\label{Fig:M_EC_Single_Low}
\end{center}
\end{figure}

In \figref{Fig:M_EC_r2_S} and \figref{Fig:M_EC_r2_W}, the lower bound ergodic capacity of FSO channel in operation under IM/DD technique is demonstrated with varying effects of pointing error, $\xi=1$ and $6.7$, for strong atmospheric turbulence and weak atmospheric turbulence, respectively.
Expectedly, as the atmospheric turbulence conditions get severe and/or as the pointing error gets severe, the ergodic capacity starts decreasing (i.e. the higher the values of $\alpha$ and $\beta$, and/or $\xi$, the higher will be the ergodic capacity). One of the most important outcomes of \figref{Fig:M_EC_r2_S} and \figref{Fig:M_EC_r2_W} are the asymptotic results for the ergodic capacity via two different methods. It can be seen that at high SNR, the asymptotic expression, via Meijer's G function expansion, derived in \eqref{Eq:MCAP} (i.e. utilizing all the terms in the summation) converges rather slowly. Based on the effects of the turbulent parameters and the pointing error, the appropriate dominant term(s) are selected and we can see that these respective dominant term(s) also converge though relatively quite faster than the case where we employ all the terms. On the other hand, the asymptotic expression, via utilizing moments, derived in \eqref{Eq:M_Moments_Derivative_n0} gives very tight asymptotic results in high SNR regime. Interestingly enough, it can be clearly seen that the two-dominant terms of \eqref{Eq:MCAP} (derived via Meijer's G function expansion) signified by the two $1$'s present in the Meijer's G function of the lower bound ergodic capacity results in \eqref{Eq:M_EC} and \eqref{Eq:M_Moments_Derivative_n0} (derived via moments) overlap. Finally, \figref{Fig:M_EC_Single_Low} presents tight asymptotic results for the ergodic capacity in low SNR regime derived in \eqref{Eq:M_EC_Single_Low}.
It is important to observe that non-intuitively, for low SNR regime (i.e. \figref{Fig:M_EC_Single_Low}), as the pointing error gets severe, the ergodic capacity starts increasing (i.e. the higher the value of $\xi$, the lower will be the ergodic capacity).

\begin{figure}[h]
\begin{center}
\includegraphics[scale=0.29]{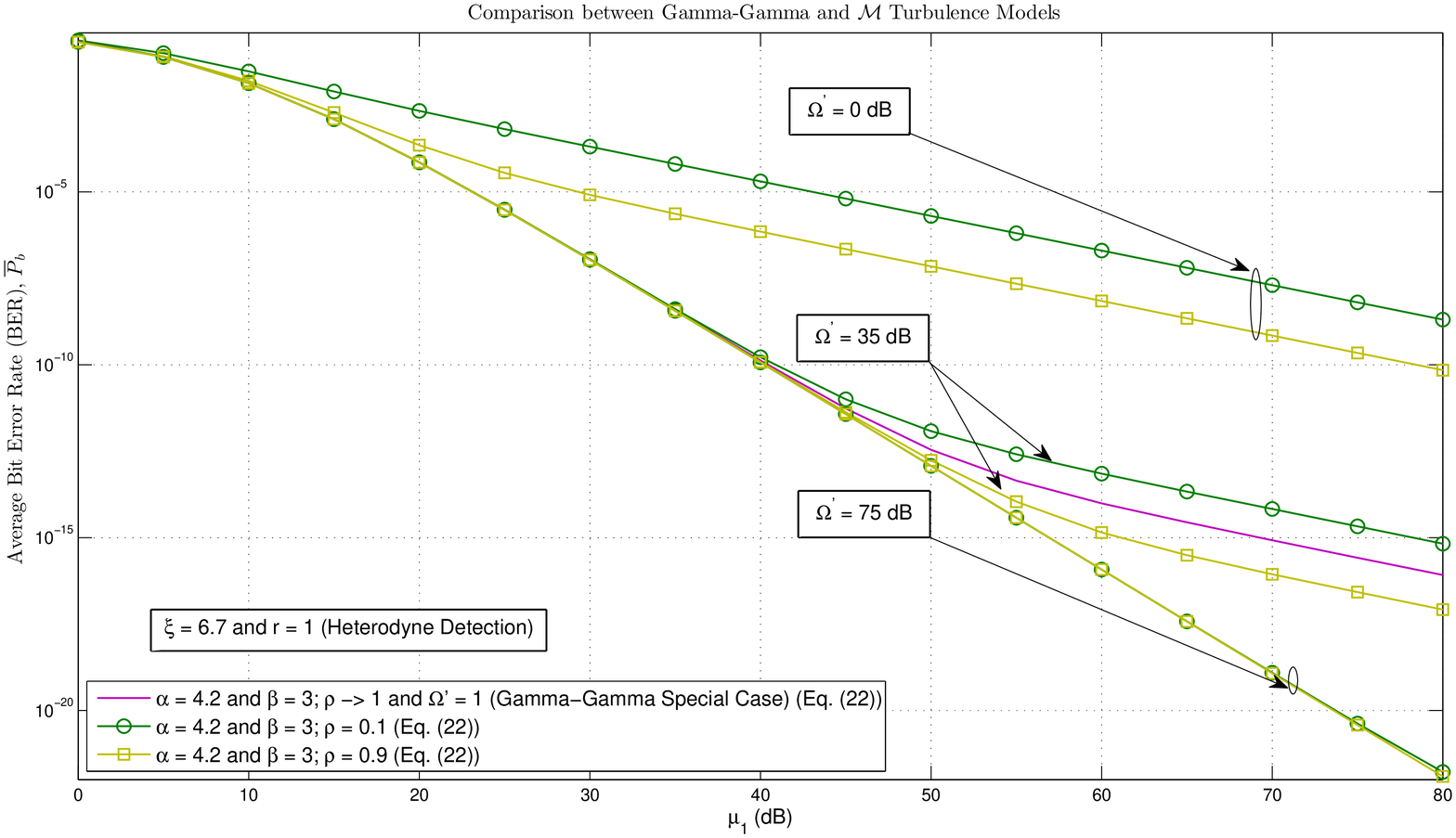}
\caption{Comparison of the FSO link performance with Gamma-Gamma and $\mathcal{M}$ turbulent channels with fixed $\alpha, \beta,$ and $\xi$.}
\label{Fig:GGvsM}
\end{center}
\end{figure}

Finally in \figref{Fig:GGvsM}, we demonstrate the relative performance of $\mathcal{M}$ turbulent channels with Gamma-Gamma turbulent channels.
It is interesting to see how $\rho$ and $\Omega^{'}$ behave. It can be observed that $\rho$ has a significant effect on the performance though as the $\Omega^{'}$ increases much beyond $60$ dB, the effect of $\rho$ nullifies. Similar trend is observed for the variations with $\Omega^{'}$ itself.

\section{Concluding Remarks}
We presented unified expressions for the PDF, the CDF, the MGF, and the moments of the average SNR of an FSO link operating over $\mathcal{M}$ turbulence. Capitalizing on these expressions, we presented new unified formulas for various performance metrics including the OP, the SI, the error rate of a variety of modulation schemes, and the ergodic capacity in terms of Meijer's G function except for the SI that was in terms of simple elementary functions. Further, we derived and presented novel asymptotic expressions for the OP, the average BER, and the ergodic capacity in terms of basic elementary functions via utilizing Meijer's G function expansion given in the Appendix and via utilizing moments too for the ergodic capacity asymptotes. In addition, this work presented simulation examples to validate and illustrate the mathematical formulation developed in this work and to show the effect of the atmospheric turbulence conditions severity and the pointing errors severity on the system performance.

These results demonstrate the unification of various FSO turbulent scenarios into a single expression allowing one to utilize this unified expression and derive the required expression for one's objective. Additionally, one can easily utilize the unified analysis over the $\mathcal{M}$ FSO turbulent channels to obtain many other turbulent channels, as per need, as its special case. Furthermore, having these unified asymptotic results opens door for simpler further analysis over more complex systems undergoing these FSO turbulent channels.


\section*{Appendix: Meijer's G Function Expansion}\label{App:AppendixA}
The Meijer's G function can be expressed, at a very high value of its argument, in terms of basic elementary functions via utilizing Meijer's G function expansion in \cite[Theorem 1.4.2, Eq. (1.4.13)]{mathai4} and $\lim_{x\rightarrow 0^{+}}\Hypergeom{c}{d}{e}{f}{x}=1$ \cite{renzo1} as
\begin{equation}\label{Eq:G-Expansion}
\begin{aligned}
\lim_{z\rightarrow \infty^{+}}&\MeijerG[right]{m,n}{p,q}{z}{a_1,\dots,a_n,\dots,a_p}{b_1,\dots,b_m,\dots,b_q}=\sum_{k=1}^{n}z^{a_k-1}\\
&\times\frac{\prod_{l=1;\,l\neq k}^{n}\Gamma(a_k-a_l)\prod_{l=1}^{m}\Gamma(1+b_l-a_k)}{\prod_{l=n+1}^{p}\Gamma(1+a_l-a_k)\prod_{l=m+1}^{q}\Gamma(a_k-b_l)},
\end{aligned}
\end{equation}
where $a_k-a_l\neq 0, \pm 1, \pm 2,\dots; (k, l=1,\dots,n; k\neq l)$ and $a_k-b_l\neq 1, 2, 3,\dots; (k=1,\dots,n; l=1,\dots,m)$.


\bibliographystyle{IEEEtran}
\bibliography{References}

\end{document}